\documentclass[acmlarge,screen,prologue,table,xcdraw,nonacm]{acmart}
\usepackage{graphicx}
\usepackage{amsmath} 
\usepackage{color}

\usepackage{url}
\usepackage{units}
\usepackage{dirtytalk}
\usepackage{subfigure}
\usepackage{algorithm}
\usepackage{algorithmicx}
\usepackage{algpseudocode}
\usepackage{soul}
\usepackage{dirtytalk}
\usepackage{appendix}
\usepackage{svg}

\normalsize

\begin{document}

\title{Can Large Language Models Be Good Companions? An LLM-Based Eyewear System with Conversational Common Ground}
\author{Zhenyu Xu}
\authornote{Equal contribution}
\email{zyxu22@m.fudan.edu.cn}
\orcid{0009-0008-4497-724X}
\author{Hailin Xu}
\authornotemark[1]
\email{hlxu22@m.fudan.edu.cn}
\orcid{0009-0000-7570-4078}
\author{Zhouyang Lu}
\authornotemark[1]
\email{zylu22@m.fudan.edu.cn}
\orcid{0009-0005-6159-6914}

\affiliation{%
  \institution{School of Computer Science, Fudan University}
  \city{Shanghai}
  \country{China}
  \postcode{200438}
}

\author{Yingying Zhao}
\email{yingyingzhao@fudan.edu.cn}
\orcid{0000-0001-5902-1306}
\affiliation{%
  \institution{School of Computer Science, Fudan University}
  \city{Shanghai}
  \country{China}
  \postcode{200438}
}

\author{Rui Zhu}
\email{rui.zhu@city.ac.uk}
\orcid{0000-0002-9944-0369}
\affiliation{%
  \institution{Bayes Business School, City, University of London}
  \city{London}
  \country{United Kingdom}
  \postcode{EC1Y 8TZ}
}

\author{Yujiang Wang}
\email{yujiang.wang@oxford-oscar.cn}
\orcid{0000-0002-6220-029X}
\affiliation{%
  \institution{Oxford Suzhou Centre for Advanced Research}
  \city{Suzhou}
  \country{China}
}

\author{Mingzhi Dong}
\email{mingzhidong@gmail.com}
\orcid{0000-0002-8897-5931}
\author{Yuhu Chang}
\authornote{Corresponding authors}
\email{yhchang@fudan.edu.cn}
\orcid{0000-0003-2607-916X}
\renewcommand{\shortauthors}{Anonymous Authors}
\affiliation{%
  \institution{School of Computer Science, Fudan University}
  \city{Shanghai}
  \country{China}
  \postcode{200438}
}

\author{Qin Lv}
\email{qin.lv@colorado.edu}
\orcid{0000-0002-9437-1376}
\affiliation{%
  \institution{Department of Computer Science, University of Colorado Boulder}
  \city{Boulder}
  \state{Colorado}
  \country{United States}
  \postcode{80309}
}

\author{Robert P. Dick}
\email{dickrp@umich.edu}
\orcid{0000-0001-5428-9530}
\affiliation{%
  \institution{Department of Electrical Engineering and Computer Science, University of Michigan}
  \city{Ann Arbor}
  \state{Michigan}
  \country{United States}
  \postcode{48109}
}

\author{Fan Yang}
\email{yangfan@fudan.edu.cn}
\orcid{0000-0003-2164-8175}
\affiliation{%
  \institution{School of Microelectronics, Fudan University}
  \city{Shanghai}
  \country{China}
  \postcode{201203}
}

\author{Tun Lu}
\email{lutun@fudan.edu.cn}
\orcid{0000-0002-6633-4826}

\author{Ning Gu}
\email{ninggu@fudan.edu.cn}
\orcid{0000-0002-2915-974X}

\author{Li Shang}
\authornotemark[2]
\email{lishang@fudan.edu.cn}
\orcid{0000-0003-3944-7531}
\affiliation{%
  \institution{School of Computer Science, Fudan University}
  \city{Shanghai}
  \country{China}
  \postcode{200438}
}

\renewcommand{\shortauthors}{Xu, Xu, and Lu, et al.}

\begin{abstract}
Developing chatbots as personal companions has long been a goal of artificial intelligence researchers. 
Recent advances in Large Language Models (LLMs) have delivered a practical solution for endowing chatbots with anthropomorphic language capabilities. 
However, it takes more than LLMs to enable chatbots that can act as companions.
Humans use their understanding of individual personalities to drive conversations. Chatbots also require this capability to enable human-like companionship. They should act based on personalized, real-time, and time-evolving knowledge of their owner.
We define such essential knowledge as the \textit{common ground} between chatbots and their owners, and we propose to build a common-ground-aware dialogue system from an LLM-based module, named \textit{OS-1}, to enable chatbot companionship.
Hosted by eyewear, OS-1 can sense the visual and audio signals the user receives and extract real-time contextual semantics. 
Those semantics are categorized and recorded to formulate historical contexts from which the user's profile is distilled and evolves over time, i.e., OS-1 gradually learns about its user. 
OS-1 combines knowledge from real-time semantics, historical contexts, and user-specific profiles to produce a common-ground-aware prompt input into the LLM module. The LLM's output is converted to audio, spoken to the wearer when appropriate.
We conduct laboratory and in-field studies to assess OS-1's ability to build common ground between the chatbot and its user. The technical feasibility and capabilities of the system are also evaluated. 
OS-1, with its common-ground awareness, can significantly improve user satisfaction and potentially lead to downstream tasks such as personal emotional support and assistance. 

\end{abstract}



\maketitle

\section{Introduction}
\label{sctn::intro}
\begin{figure}[htbp]
    \centering
    \includegraphics[scale=0.75]{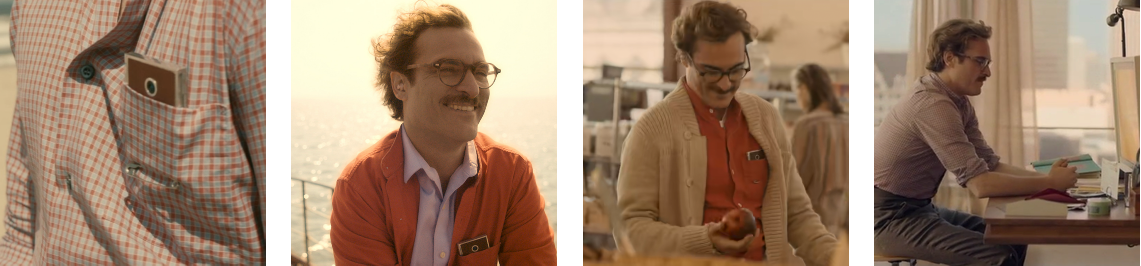}
    \caption{Examples of Samantha (the red, rectangular smartphone in his pocket) accompanying Theodore in his daily life. }
    \label{Samantha_cases}
\end{figure}

It has long been a vision for chatbots to be personal, human-like companions. One classic example is Samantha, a portable artificial intelligence dialogue system depicted in the movie ``Her'' (see Figure~\ref{Samantha_cases}). The protagonist, Theodore, carries Samantha in his chest pocket. She has a camera and microphone. Samantha sees what Theodore sees, hears what he hears, and chats with him using an earbud. She shares his joys and sorrows during work and leisure. Through day-by-day interactions, Samantha gradually learns Theodore's personality, preferences, and habits. She offers companionship, emotional support, and assistance, and eventually becomes a nearly indispensable part of Theodore's life.

To realize this vision, several technical challenges must be addressed. The limited linguistic and cognitive capabilities of natural language processing (NLP) have been recognized as major barriers to personalized dialogues~\cite{adamopoulou2020chatbots}. 
Recent advances in large language models (LLMs) such as ChatGPT (based on the GPT-3.5 LLM model)~\cite{chatgpt} and GPT-4~\cite{gpt4report} have largely removed this barrier and opened the possibility of supporting natural and human-like conversations.
Pre-trained on massive amounts of text data, LLMs have the ability to encode a vast amount of world knowledge. These capabilities allow LLMs to generate coherent and diverse responses; this is crucial for natural conversation. Additionally, through supervised instruction fine-tuning and reinforcement learning with human feedback~\cite{ouyang2022training}, LLMs can be adapted to follow human instructions while avoiding creating harmful or inappropriate content.
Inspired by the powerful language modeling capabilities of LLMs, the question arises: \emph{\say{Can LLM-based chatbots serve as personal companions in daily life?}} 

We argue that the answer today remains, \say{Not without further capabilities}. Despite impressive human-like language capabilities, LLMs lack \textbf{\emph{common ground}, preventing LLM-based chatbots from being personal companions}.
Based on research in linguistics~\cite{clark1996using}, psychology~\cite{GUYDISH2021100877}, and Human-Computer Interaction (HCI)~\cite{clark2019makes}, having common ground is essential for successful and meaningful conversations. This common ground can stem from shared personal experiences, interests, and other factors. For example, when initiating a dialogue with other people, we either ask questions to establish common ground or presuppose certain common ground already exists~\cite{skantze2023open}.  
\textbf{It is challenging for an LLM to establish a mutual understanding with a person.} The common ground between humans is usually implicit and subjective~\cite{zhou2022reflect,CLARK1989259}. Therefore, it is not practical to expect users to provide common ground information explicitly. Also, LLMs are generally not equipped to perceive a user's context, e.g., their physical surroundings or daily experiences. Without such personal context, LLMs struggle to comprehend a user's visual surroundings, speech, daily events, and behaviour (e.g., personality traits~\cite{allport1927concepts}, habits). This prevents them from establishing common ground with their users. 

This work is inspired by the powerful language generation capabilities of LLMs~\cite{chatgpt,gpt4report,chang2023survey,beltagy2019scibert} and motivated by their lack of personal context awareness necessary to establish common ground. To bridge these gaps, we formulate the following research questions (RS).

\emph{RS1}. Does personal context help LLM-based dialog systems establish common ground with their users?

\emph{RS2}. In what ways do different types of personal context contribute to personalized LLM-based dialog system responses?

We argue that the answer today is, \say{Ubiquitous personal context helps establish common ground between LLM-based dialogue systems and their users, and furthermore, it enables more personalized responses}. 
To verify our hypotheses, this work breaks \emph{personal context} into the following three categories in the temporal dimension, and describes the design of an LLM-based smart eyewear system to achieve ubiquitous personal context capturing and use. 

\begin{itemize}

\item \textbf{Real-time context} refers to momentary semantics inferred from the user's ongoing speech and visual surroundings. These semantics help LLMs understand the meanings of the user's speech and visual perceptions, enabling the generation of appropriate responses. 

\item \textbf{Historical context} is a summary of the past real-time context time series. It organizes the user's daily events (e.g., activities) and dialogue contents by clustering the real-time contexts into temporal units. This information helps LLMs maintain the coherence and continuity of the dialogue, and enables it to avoid repeating or contradicting previous statements. 

\item \textbf{User profiles} are distilled historical information related to the user's personality, habits, and preferences, which are revealed during interaction with the dialogue system. They can enable LLMs to incorporate additional human-like qualities by adapting to the user's personality and long-term goals, resulting in more consistent and anthropomorphic responses.
\end{itemize}

\begin{figure}[htbp]
    \centering
    \includegraphics[width=0.6\textwidth]{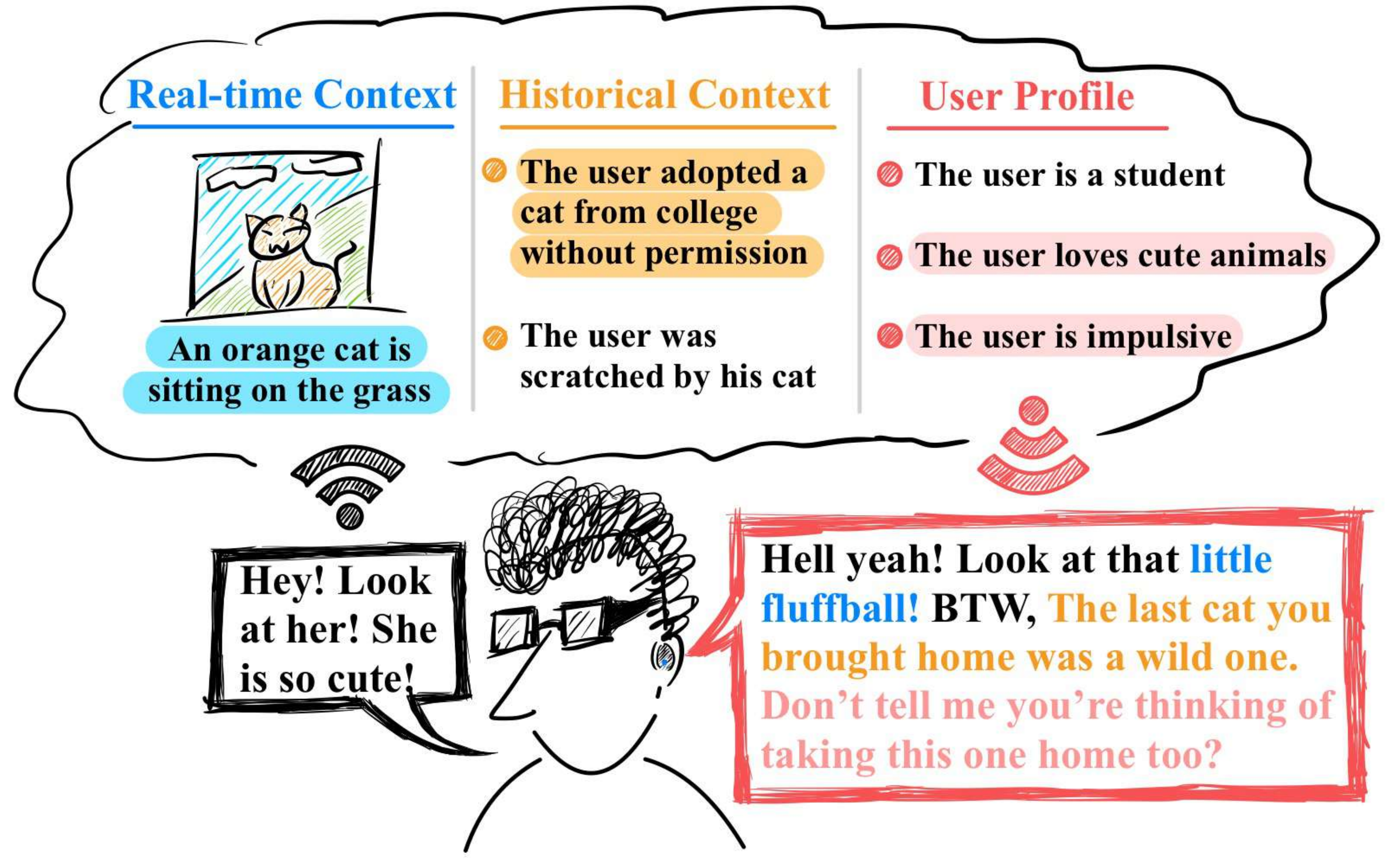}
    \caption{Conceptual overview of OS-1 workflow. }
    \label{system_workflow}
\end{figure}

To answer the two RS above, this work presents OS-1\footnote{``OS-1'' was the name for Samantha's underlying implementation
in the movie ``Her".}, 
the first LLM-based chatbot system aware of the conversational common ground with its users. 
Residing on smart glasses, OS-1 captures the visual and audio signals received by its user as input, builds conversational common ground gradually, and generates personalized dialogues at proper times. As portrayed in Figure \ref{system_workflow}, OS-1 consists of four central modules.

\begin{enumerate}

    \item \textbf{Real-time context capture.} The smart eyewear first perceives the user's in-situ visual and audio signals through the built-in camera and microphone, and transfers them to the cloud on the fly. These two types of information are essential to understanding the user's ongoing status. Using a vision-language model (e.g., LLaVA~\cite{liu2023llava}) and a speech recognition model (e.g., Whisper~\cite{radford2023robust}) deployed in the cloud, we can infer the semantic description of images and transcribe voice data into text. Also, by combining these two data modalities, OS-1 can infer the user's current activity, location, and other information inferred from the user's surroundings. These forms of information constitute the real-time context.
		
    \item \textbf{Historical context extraction}. To ensure long-term coherence and consistency in dialogues, it is important to remember the user's historical information.
    To this end, we design a clustering method that extracts the relevant information, such as daily events, from the accumulated real-time contexts, thus forming the historical context. The clustering method removes redundancy between inter-real-time contexts and produces event-level descriptions that are easy to summarize. We designed indexing methods along temporal, spatial, and semantic dimensions to facilitate efficient retrieval of historical contexts from different perspectives.
    
    \item \textbf{User profile distillation}. To better understand users' profiles, we analyze their historical context to form a user profile that includes their personality, preferences, and life habits. However, our inference of the user profile may be biased or contain errors due to limited interactions. To address this issue, we design an update scheme that can revise the current user profile based on the historical context and past user profiles.

    \item \textbf{Personalized response generation}. 
When the user launches a conversation with OS-1, a real-time context will be extracted from the eyewear's video and audio stream, and OS-1 will retrieve the update-to-date knowledge database of the user, namely the historical context and user profile, using the latest real-time context as the key.
This retrieval process is achieved by the strategy of multi-LLM agents~\cite{zhao2023survey,xi2023rise}.
The resulting personal context, which contains personality information sufficient enough to drive common-ground-aware conversations, is used as the LLM prompt to generate an appropriate response.
This textual response is converted into audio delivered to the glasses wearer, ending this conversation cycle and waiting for the next one.
    
    
    

\end{enumerate}
We conduct in-lab experiments and in-field pilot studies to evaluate OS-1's ability to establish common ground using the captured and refined personal contexts. The ability to do these things would enable OS-1 to facilitate better conversations that satisfy its users. 
Inspired by the idea of using human-like features as measurements of conversational response quality~\cite{1905.04071}, this work proposes customized human evaluation metrics to evaluate the performance of OS-1.
More specifically, we first propose a human evaluation metric dubbed \textit{Grounding} score to evaluate how well OS-1 can build up common ground with its users. Also, we propose three more fine-grained metrics -- Relevance, Personalization, and Engagement score, to evaluate the relevance of the system's responses to the real-time context, the relationship between the responses and the user's historical and profile context, as well as the level of interest a user shows in the response. Laboratory results show that, compared to the baseline method without any personal contexts, OS-1 improves the Grounding score by 42.26\%.
Also, OS-1 substantially improves the performance by 8.63\%, 40.00\%, and 29.81\% in Relevance, Personalization, and Engagement score, respectively. The in-field pilot study further shows that the Grounding score exhibits an increasing trend over time, which indicates that OS-1 is capable of continuously reaching common ground with users through long-term interactions. We also explain its behavior into potential applications: emotional support, and personal assistance, and conduct semi-structured interviews to provide qualitative insights.

Our work makes the following contributions.
\begin{enumerate}

\item  We present a novel concept of personal context and a human evaluation metric Ground score to assess the ability of an LLM-based dialogue system to reach mutual understanding. This opens up the possibility of supporting personal companionship applications.

\item We design and implement an always-available smart eyewear LLM based personal dialogue system that captures the user's multi-modal surroundings on-the-fly, generates personal context, and engages in personalized conversation with the user. One of the greatest strengths of the system is its ability to achieve the above automatically without introducing any additional cognitive load or interaction requirements on users, thereby enhancing the user experience under various HCI scenarios.

\item We propose a novel method to capture, accumulate, and refine the personal context from user multi-modal contexts and dialogue histories, and a multi-dimensional indexing and retrieval mechanism that integrates multiple personal contexts to enable personalized responses. Our method facilitates dynamic adaptation to the user's surroundings, experiences, and traits, enabling an engaging and customized conversation experience. 

\item We conducted an in-lab study and a pilot study to evaluate the impact of using personal context within the dialogue system. Our results show the superior performance of the proposed system in gradually reaching a better mutual understanding over time.

\end{enumerate}

\section{Related Work}
\label{sec:related_work}
In this section, we provide a brief overview of related work in the following areas: (1) large language models, (2) multimodal dialogue systems, (3) personalized dialogue systems, and (4) wearable dialogue systems.

\subsection{Large Language Models}
Large language models are recent innovations that revolutionized the field of natural language processing (NLP) and influenced other areas. 
LLMs are pre-trained on large-scale corpora. Models such as GPT-3.5~\cite{chatgpt}, GPT-4~\cite{gpt4report}, Vicuna~\cite{vicuna2023}, Llama 2~\cite{touvron2023llama}, Qwen~\cite{qwen} and Falcon~\cite{falcon}, 
have demonstrated impressive language understanding and modelling capabilities unseen in neural networks of smaller parametric scales~\cite{zhao2023survey}. 
In addition to outstanding language intelligence, LLMs also have surprising and valuable capabilities rare in, or absent from, smaller-scale DNNs, a phenomenon called emergent capabilities~\cite{wei2022emergent}.
One such capability is in-context learning~\cite{dong2022survey}, in which the LLMs need only be exposed to a few examples for its learning to be transfer to a new task/domain.
Additionally, through supervised instruction fine-tuning and reinforcement learning with human feedback (RLHF)~\cite{ouyang2022training}, LLMs can follow human instructions. This feature has enabled LLMs to contribute to a variety of tasks~\cite{bubeck2023sparks} such as text summarization~\cite{pu2023summarization} and sentiment analysis~\cite{wang2023chatgpt}.

The Chain-of-Thought (CoT) method~\cite{wei2022chain}, on the other hand, has shown that LLMs can be guided to conduct complex reasonings by prompting to generate intermediate steps. Similarly, for the complex reasoning task~\cite{chu2023survey}, works on X-of-Thought (XoT) move away from CoT's sequential, step-by-step thought chain and structure reasoning in a non-linear manner, such as Tree-of-Thoughts (ToT)~\cite{yao2023tree} and Graph-of-Thoughts (GoT)~\cite{besta2023graph}. LLM-based agents are also attracting researchers' attention. ReAct~\cite{yao2022react} generates thoughts and actions in an interleaved manner, leading to human-like decisions in interactive environments. In the planning-execution-refinement paradigm~\cite{zhao2023survey}, AutoGPT~\cite{autogpt} follows an iterative process reminiscent of human-like problem-solving, i.e., a plan is proposed, executed, and then refined based on feedback and outcomes. Systems like Generative Agents~\cite{park2023generative} and ChatDev~\cite{qian2023communicative} explore multi-agent collaboration; agents interact with the environment and exchange information with each other to collaborate and share task-relevant information.

In this work, we generally follow the prompt generation paradigms in ICL and CoT~\cite{wei2022chain}. Our work is also inspired by the planning-execution-refinement paradigm~\cite{zhao2023survey}; that is, just like agents, our system investigates the context to generate a plan that is used to select and action. The plan is iteratively refined based on user feedback when creating a dialogue strategy.

\subsection{Multimodal Dialogue Systems}
Multimodal dialogue systems can leverage contextual information from multiple modalities, such as text and images, to improve users' experience. The visual dialogue task was first introduced by Das et al.~\cite{Das_2017_CVPR}, involving two participants in an image-based question-answering task, where a person asks a question about an image and a chatbot gives a response. Mustafazade et al.~\cite{mostafazadeh2017imagegrounded} introduced the image-grounded conversation (IGC) task, which improves the conversation experience by allowing the system to answer and ask questions based on visual content. Despite progress in extending dialogue context modalities, these early works do not use natural language modelling capabilities.

Recently, multimodal dialogue systems used the capabilities of both the visual and language models. These vision-language models (VLMs)~\cite{li2023blip2, ye2023mplugowl,liu2023llava, zhu2023minigpt4} can generate coherent language responses consistent with the visual context. However, they still face challenges in generating natural dialogues that occur in real-life interactions. Furthermore, Li et al.~\cite{li2023mimicit} introduced the interactive vision-language task MIMIC-IT, which allows dialogue systems to engage in immersive conversations based on the multimodal context.

In this work, we combine two state-of-the-art models, i.e., the visual understanding capabilities of VLMs with the dialogue capabilities of LLMs, to enhance the conversational experience.

\subsection{Personalized Dialogue Systems}
In the dialogue system research field, user profiles such as personality, preferences, and habits can be extracted from user interactions to support personalized dialogue~\cite{zhang2018personalizing}. However, previous studies mainly focus on short-term dialogues, not gradually increasing their understanding of users via long-term interactions. Recently, Xu et al.~\cite{xu-etal-2022-beyond} proposed a long-term dialogue task including user profiles. However, this task does not consider the key elements of extracting, updating, and utilizing user profiles. To address this limitation, Xu et al.~\cite{xu2022long} recently proposed to identify user personas from utterances in a conversation, which are then used to generate role-based responses.

More recently, Ahn et al.~\cite{ahn2023mpchat} proposed to incorporate visual modalities to enhance the understanding of user profiles from recorded episodic memory. It overcomes the limitation of relying on text-only conversations. However, these episodic memories mainly consist of images and texts shared on social media rather than users' real-life experiences. Combining episodic memory with user profiles, Zhu et al.~\cite{zhong2023memorybank} utilized LLMs to summarize conversations into episodic memories and user profiles, which are then stored in a vector database and retrieved based on the dialogue context in subsequent conversations, resulting in personalized responses.



In this work, we generate historical context and user profile from multimodal information captured in real-world scenarios. Compared with previous literature, our work utilizes more real-time user information sources. Furthermore, we introduce a mechanism for accumulating user information, enabling the system to enhance its knowledge of users over time.

\subsection{Wearable Dialogue Systems}
Wearable dialogue systems are a developing area of research that combines wearable technology with conversational AI. Currently, wearable dialogue systems focus on specific user groups or application domains, such as the visually impaired or the healthcare domain. Chen et al.~\cite{9258014} proposed a wearable dialogue system for visually impaired individuals that employs smart eyewear with 3D vision, a microphone, and a speaker to facilitate outdoor navigation through conversation. Ozono et al.~\cite{ozono2022encouraging} proposed a system that combines wearable devices and interactive agents, mainly aimed at promoting and encouraging elderly people to take better care of their health. The approach involves integrating health data into conversations with users, to make elderly people aware of their health issues and encourage self-care.

Shoji et al.~\cite{9427763} proposed a dialogue system based on smart eyewear, which can interact with users through voice and provide daily life information, such as weather. Additionally, it also gathers users' biometric data, such as pulse and body temperature, to offer health management guidance through conversation. Kocielnik et al.~\cite{10.1145/3214273} proposed a mobile dialogue system that collects physical activity data through fitness trackers and guides users to reflect on their daily physical activities through conversations. Calvaresi et al.~\cite{10.1145/3486622.3493992} proposed a mobile health assistant that monitors diet and offers suggestions through conversations. It can track nutritional information by scanning product barcodes or analyzing food images, offer dietary recommendations, and utilize the user's GPS location to recommend nearby restaurants.

In contrast with prior work, our work primarily aims to offer personalized conversations and companionship to the user. By combining wearable technology with advanced conversational AI, our goal is to build a seamless and natural interaction experience that goes beyond functional support. It incorporates contextual information to continually improve the quality of the interaction and adapt to the user's experiences and preferences over time, thereby creating a sustainable personal companion.

\section{System Design}
\label{sec:system_design}

This section first provides an overview of the proposed ubiquitous context-aware dialogue system OS-1. It then details the four core modules of OS-1. It then describes the system implementation process.

\subsection{Overview}

During the design of OS-1, we consider the following five aspects of requirements: 1) episodic understanding, 2) memorization ability, 3) personalization awareness, 4) personalized responsiveness, and 5) ubiquitous accessibility.


\paragraph{\textbf{Episodic Understanding}} To achieve episodic understanding, OS-1 needs to perceive the user's ongoing conversation and understand the in-situ context in real-time, including the visual and auditory surroundings, location, and activity. Therefore, we equip the smart eyewear-based system with cameras, microphones, and speakers to capture the surrounding images and speech, which are converted into text using the vision-language model, LLaVA~\cite{liu2023llava}, and the speech recognition model,  Whisper~\cite{radford2023robust}. The converted texts from the images and speech are then fused to form a prompt. Next, we utilize the responses of GPT-3.5 to infer the user's real-time context via the prompt. 

\paragraph{\textbf{Memorization Ability}}
To enable memorization, OS-1 generates, stores, and recalls the historical contexts, including the user's past daily events and dialogue content. To reduce the redundant storage of past real-time contexts and achieve effective retrieval, OS-1 summarizes the past real-time contexts via a clustering approach that considers semantic similarity. Highly similar real-time contexts are clustered and summarized into distinct events using GPT-3.5, thus serving as historical contexts. Additionally, we propose a mechanism to generate the temporal, spatial, and semantic indices for the historical contexts, which are stored in the vector database, Milvus~\cite{2021milvus}, enabling retrieval of similar historical contexts in these three dimensions.

\paragraph{\textbf{Personalization Awareness}}
To further enhance the personalization of OS-1, we propose to distill and update user profiles over time based on inference of the user's personality, preferences, social background, and life habits from the historical contexts via GPT-3.5. Our updating mechanism assigns a confidence score to each user profile to guide the review and revision of existing profiles. When a new user profile is generated, OS-1 retrieves the most semantically similar existing profile from Milvus~\cite{2021milvus}. The new and existing profiles are merged to construct a prompt for GPT-3.5, which generates an updated user profile and is then stored in Milvus~\cite{2021milvus}. 

\paragraph{\textbf{Personalized Responsiveness}}
To generate personalized responses, we design two LLM-based agents: the dialogue strategy agent and the information retrieval agent. The dialogue strategy agent decides the conversational strategy, while the information retrieval agent aims at retrieving the relevant information from historical contexts and user profiles following the planned strategy. The personal context, including the retrieved information and real-time context along with the dialogue strategy, is used to construct a prompt for GPT-4 to generate personalized responses.



\paragraph{\textbf{Ubiquitous Accessibility}}
To enable conversation anytime and anywhere, a lightweight, portable, battery-powered hardware device is required, implying constraints on computing ability and battery capacity. The challenge lies in developing a system with the above four capabilities in the presence of these constraints. 
Therefore, our architecture assigns basic functions, including image capture, audio recording, and audio playback, to the smart eyewear device while offloading more compute- and energy-intensive functions, including real-time context capture, historical context extraction, user profile distillation, and personalized response generation to the cloud.

\subsection{Workflow} \label{subsec:cloud-workflow}

\subsubsection{Overall Pipeline}
\begin{figure}[htbp]
    \centering
    \includegraphics[width=1\textwidth]{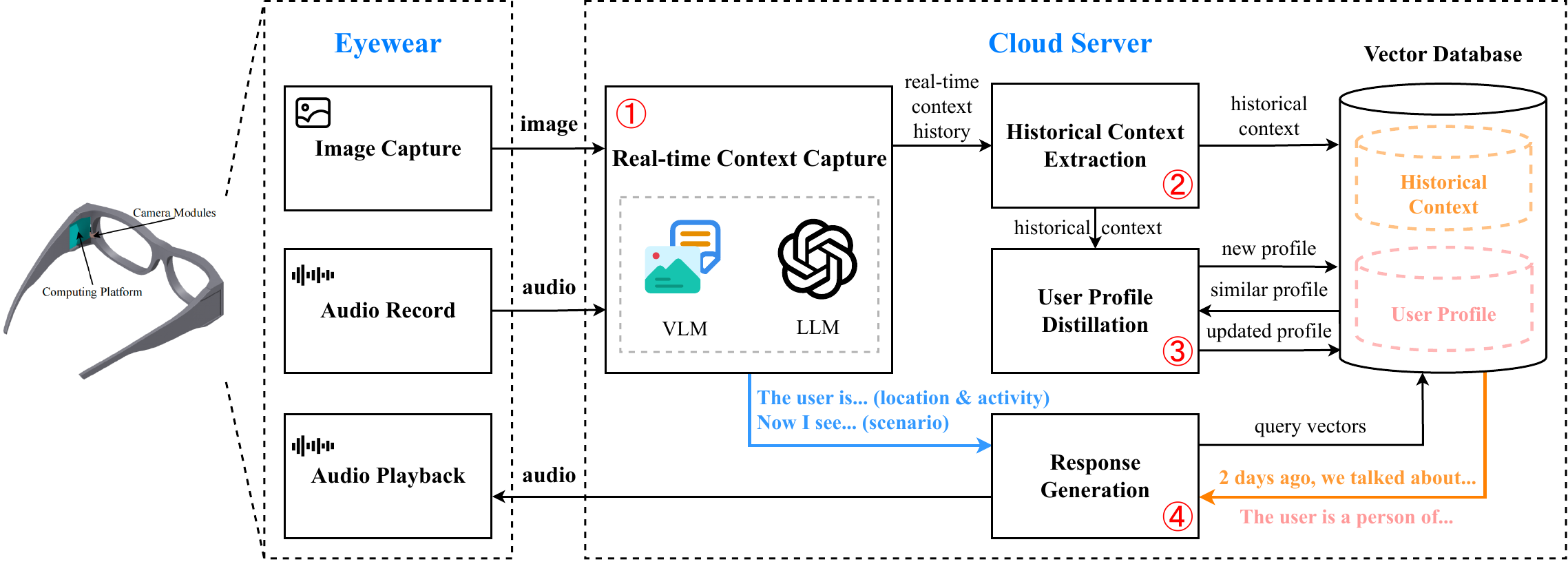}
    \caption{The overall framework of OS-1. The image and audio captured by the eyewear are sent to a cloud server for processing with four sequential steps. OS-1 prepares a response that is converted to audio on the eyewear. 
    }
    \label{fig:system_framework}
\end{figure}

The overall framework is shown in Figure~\ref{fig:system_framework}.
In the real-time context capture stage, the eyewear obtains the surrounding image and audio and transmits them to the cloud for real-time context capture. In the historical context extraction stage, daily events and conversation summaries are extracted from the history of real-time contexts, which are then assigned multi-dimensional indices and an importance score, then stored in the vector database as historical context. In the user profile distillation stage, OS-1 generates a new user profile from historical contexts and retrieves a similar user profile from the vector database. The new user profile and the similar user profile are merged to obtain an updated user profile, which is stored in the vector database. In the personalized response generation stage, the dialogue strategy agent plans the conversational strategy, while the information retrieval agent retrieves the relevant information from historical contexts and user profiles. OS-1 then combines the real-time context and the retrieved information with the dialogue strategy to generate text responses using GPT-4. Subsequently, these responses are converted to speech and played back on the eyewear. During implementation, we use both GPT-3.5 and GPT-4. We select GPT-4 to generate the final response for its superior quality and GPT-3.5 for other tasks to control the overall cost of our project. Hereafter, GPT-3.5 is referred to as LLM-Base, while GPT-4 is referred to as LLM-Large.


\subsubsection{Real-time Context Capture} \label{subsubsec:episodic-context-capture}
OS-1 captures real-time visual and audio signals through the built-in camera and microphone on the smart glasses. It then uses a vision-language model to convert visual signals into descriptions, providing textual descriptions of scenes, such as ``a desk with a laptop''. Additionally, an audio speech recognition model also transcribes audio signals into text, recognizing what the user said, such as ``I am so busy''. By semantically combining the textual descriptions from visual and audio signals, OS-1 leverages the knowledge of LLM-Base to infer the user's location and activity. For example, it may determine that the user is in the ``office'' and the user's activity is ``working''. The texts obtained from the image and audio signals and the location and activity inferred by LLM-Base form the real-time context, which assists OS-1 to understand the user's current situation.



Specifically, during a conversation between the user and OS-1, the audio signal corresponding to the user's $t$-th utterance is denoted as $A^t$, and the most recently captured image signal is denoted as $I^t$. OS-1 employs the speech recognition model $\mathcal N_{asr}$~\cite{radford2023robust} to transcribe $A^t$ into text, resulting in $u^t=\mathcal N_{asr}(A^t)$, where $u^t$ represents the transcribed text of the user's $t$-th utterance $A^t$. For the image signal, OS-1 employs the vision-language model $\mathcal N_{vlm}$~\cite{liu2023llava} to generate a textual description of the scene, resulting in $v^t=\mathcal N_{vlm}(I^t)$, where $v^t$ represents the caption of the image signal $I^t$. Then, we construct a prompt for LLM-Base to infer the current location and activity, $\{l^{t}, a^{t}\}=\mathcal N_{llm}(v^t, u^t)$, where $\mathcal N_{llm}$ is LLM-Base, $l^{t}$ represents the location and $a^{t}$ represents the activity. Finally, we obtain the real-time context for the $t$-th utterance, denoted as $C_{e}^{t} = \{v^{t}, u^{t}, l^{t}, a^{t}\}$.

\subsubsection{Historical Context Extraction} \label{subsubsec:historical-context-extraction}
As time goes by, OS-1 will accumulate an increasing number of real-time contexts, some of which will be largely redundant. For example, for a user who spends a long time working on a computer, the real-time context about location and activity collected by OS-1 would become repetitive. We aim to remove uninformative redundancy from stored contexts. The extracted historical context falls into two classes: daily events and conversation summaries. Daily events are triplets consisting of time, location, and activity. These allow OS-1 to store historical schedules, e.g., ``<2023-11-01 16:00:00 - 2023-11-01 17:00:00, at the gym, playing badminton>''. The conversation summary includes the topics and details of past conversations, such as ``the user mentions writing a paper and asks for tips on how to write it well''.



We propose an event clustering method that groups sequences of events into appropriate clusters and summarizes them in event-level text descriptions. To extract conversation summaries, we divide the conversation history into sessions based on contiguous time intervals. For each session, we then construct a prompt and use the summarization capability of LLM-Base to extract its summary. Furthermore, to enhance the storage and retrieval of historical contexts in the vector database, we propose an indexing mechanism that organizes the historical context into temporal, spatial, and semantic dimensions, following the format humans typically use to describe historical contexts. Additionally, it also assigns different importance scores to the historical contexts based on emotional arousal levels. The historical context with a higher arousal level is considered more important and is more likely to be referenced in subsequent conversations, as users are more likely to remember events with stronger emotional impact. We now describe the event clustering, conversation summary, and indexing mechanism.

\makeatletter
\newenvironment{breakablealgorithm}
{
		\begin{center}
			\refstepcounter{algorithm}
			\hrule height.8pt depth0pt \kern2pt
			\renewcommand{\caption}[2][\relax]{
				{\raggedright\textbf{\ALG@name~\thealgorithm} ##2\par}%
				\ifx\relax##1\relax 
				\addcontentsline{loa}{algorithm}{\protect\numberline{\thealgorithm}##2}%
				\else 
				\addcontentsline{loa}{algorithm}{\protect\numberline{\thealgorithm}##1}%
				\fi
				\kern2pt\hrule\kern2pt
			}
		}{
		\kern2pt\hrule\relax
	\end{center}
}
\makeatother

\paragraph{\textbf{Event Clustering}}
During a day, OS-1 captures a sequence of $m$ real-time contexts. For each real-time context, we use an embedding model $\mathcal N_{embed}$~\cite{reimers-2019-sentence-bert} to generate a representation vector $e^t$, denoted as $e^t=\mathcal N_{embed}(\{l^{t}, a^{t}\})$, where $\{l^{t}, a^{t}\}$ represents concatenated text descriptions of location and activity. These embedded vectors form an embedding matrix $M_e$, with each vector being a row in the matrix. Subsequently, we calculate the cosine similarity between the representation vectors of each pair of real-time contexts in the sequence to generate the similarity matrix $M_s = M_e M^T_e$. Afterward, we set a similarity threshold, which will be used to group together real-time contexts that have a cosine similarity above the threshold into an event. Due to the spatiotemporal locality of events, semantically similar real-time contexts are usually contiguous subsequences. Therefore, by sequentially traversing the overall real-time context sequence and comparing similarity with the threshold, the longest contiguous subsequence that satisfies all the following conditions is selected to cluster an event: 1) the similarity between the first element of the subsequence and the previous subsequence is below the threshold, 2) the similarities among all elements within the subsequence are above the threshold, and 3) the similarity between the last element of the subsequence and the subsequent subsequence is below the threshold. We create a prompt that summarizes a collection of real-time contexts that have been grouped together into an event; see Figure~\ref{fig:event_summary} for an illustration. Consequently, the corresponding real-time contexts for each longest subsequence are employed as parts of the prompt for LLM-Base. Finally, a summary of the event is extracted, denoted as $\{E^1,\dots, E^p\}=f_{cluster}(\{e^1,\dots, e^m\})$, where $E^i$ represents a daily event, $m$ represents the number of original unclustered real-time contexts, and $p$ represents the number of distinct events without redundancy after clustering.

\paragraph{\textbf{Conversation Summary}}
\sloppy To extract summaries from the conversation history, we set an interval threshold that determines the maximum allowed time interval within a conversation. The threshold serves as a boundary to separate conversations that exceed the interval threshold into different sessions, denoted as $\{D^1,\dots, D^q\}=f_{session}(\{u^1, b^1, \dots, u^n, b^n\})$, where $D^j$ refers to a session, $u^i$ represents the user's utterance, and $b^i$ represents the OS-1's response. After partitioning the conversation history, we construct a prompt for each session to summarize topics and details by leveraging the summarization capability of LLM-Base, denoted as $\{T^1,\dots, T^q\}=\mathcal N_{llm}(\{D^1,\dots, D^q\})$, where $T^j$ represents a conversation summary. Finally, the collections of daily event $E^i$ and conversation summary $T^j$ together form the historical context, formally represented as: $C_{h}^{1:p+q} = \{E^1,\dots, E^i,\dots, E^p, T^1, \dots, T^j, \dots, T^q\}$.

\paragraph{\textbf{Indexing Mechanism}}
We propose an indexing mechanism that organizes historical context in three dimensions: temporal, spatial, and semantic. The indexing mechanism aims to generate a list of indexing keys for textual descriptions of historical context, including daily events and conversation summaries. For example, if the historical context is ``I plan to have a picnic in the park this weekend'', the resulting indexing keys could include ``weekend plan'', ``in the park'', and ``have a picnic''. By allowing multiple indexing keys to be associated with each historical context, OS-1 can do associative retrieval in different dimensions. Specifically, we design a prompt for LLM-Base to extract the textual descriptions related to the temporal, spatial, and semantic aspects of the historical context. These extracted descriptions serve as indexing keys for the historical context. The process of generating indexing keys $K^i$ is denoted as $f_{index}$.

We also propose to incorporate emotional factors in historical context indexing. To achieve this, we design a prompt and leverage LLM-Base to evaluate the level of emotional arousal associated with a given historical context. This level determines the significance of the historical context, which is represented by an importance score ranging from 1 to 10. We assign higher importance scores to historical contexts with intensified emotional arousal, thereby increasing the likelihood of mentioning them in the conversation. This process of assigning importance scores $S^i$ is denoted as $f_{score}$.

In summary, the indexing mechanism for historical context can be formally described as follows:
\begin{enumerate}

\item generate indexing keys from multiple dimensions for each historical context, denoted as $K^i=f_{index}(C_{h}^{i})$; 

\item assign an importance score to each historical context, denoted as $S^i=f_{score}(C_{h}^{i})$; and

\item store the historical context in the vector database, along with the corresponding indexing keys and importance score.

\end{enumerate}


\subsubsection{User Profile Distillation} \label{subsubsec:user-profile-distillation}
Historical context represents the user's daily events and conversation summaries. It can therefore provide important clues about the user profile, including personality, preferences, social background, and life habits. By summarizing patterns from the historical context, OS-1 can distill the user profile and thereby improve the personalized user experience. For example, if a user frequently enjoys eating spicy food, it becomes evident that the user has a preference for spicy food. The user profile consists of a textual description of a specific aspect of the user, along with a confidence score that indicates the reliability of the information. We introduce an additional confidence score because user profile distillation is an ongoing process that aims to tackle biases and errors when inferring user profiles.

Specifically, we divide the distillation process into the following steps (see Figure~\ref{fig:user_profile}). First, we design a prompt for LLM-Base to summarize a historical context into a proposal of user profile, $C_{u}^i=\mathcal N_{llm}(C_{h}^i)$. Second, we use an embedding model~\cite{reimers-2019-sentence-bert} to encode $C_{u}^i$, resulting in a query vector. The query vector is used to retreive from the vector database the user profile with the highest cosine similarity and more than the similarity threshold, denoted as $C_u^{i'}=f_{retrieve}(C_{u}^{i})$, where $C_u^{i'}$ represents the existing user profile. If no user profile exceeds the similarity threshold, the user profile proposal is stored in the vector database. Otherwise, we design a prompt for LLM-Base to revise the concatenation of the existing user profile and the user profile proposal, denoted as $C_u^{i^*}=\mathcal N_{llm}(C_{u}^{i}, C_u^{i'})$, where $C_u^{i^*}$ represents the updated user profile. Finally, $C_u^{i^*}$ replaces $C_u^{i'}$ in the vector database. The updating mechanism enables OS-1 to rectify inaccurate user profiles and reinforce correct user profiles over time.

\subsubsection{Personalized Response Generation} 
\label{subsubsec:personalized-response-generation}

To enhance user engagement, we propose two agents, the dialogue strategy agent and the information retrieval agent, to assist in generating personalized responses. The dialogue strategy agent is responsible for planning the direction of the conversation based on real-time context and guiding users to express their opinions by asking questions, or provide additional information to drive the conversation forward. Subsequently, the information retrieval agent determines which user information to retrieve based on the dialogue strategy suggested by the dialogue strategy agent and summarizes the retrieved user information. It leverages real-time context and retrieves relevant information from historical contexts and user profiles, such as experiences and preferences. Then, OS-1 combines the real-time context and the information retrieved by the information retrieval agent as personal context, along with the dialogue strategy planned by the dialogue strategy agent, to serve as prompts for LLM-Large to generate text responses. Finally, the generated reply is converted into speech using a text-to-speech service~\cite{alibabacloudIntelligentSpeech} and transmitted to the smart eyewear device for playback.



\begin{figure}[htbp]
    \centering
    \includegraphics[width=1\textwidth]{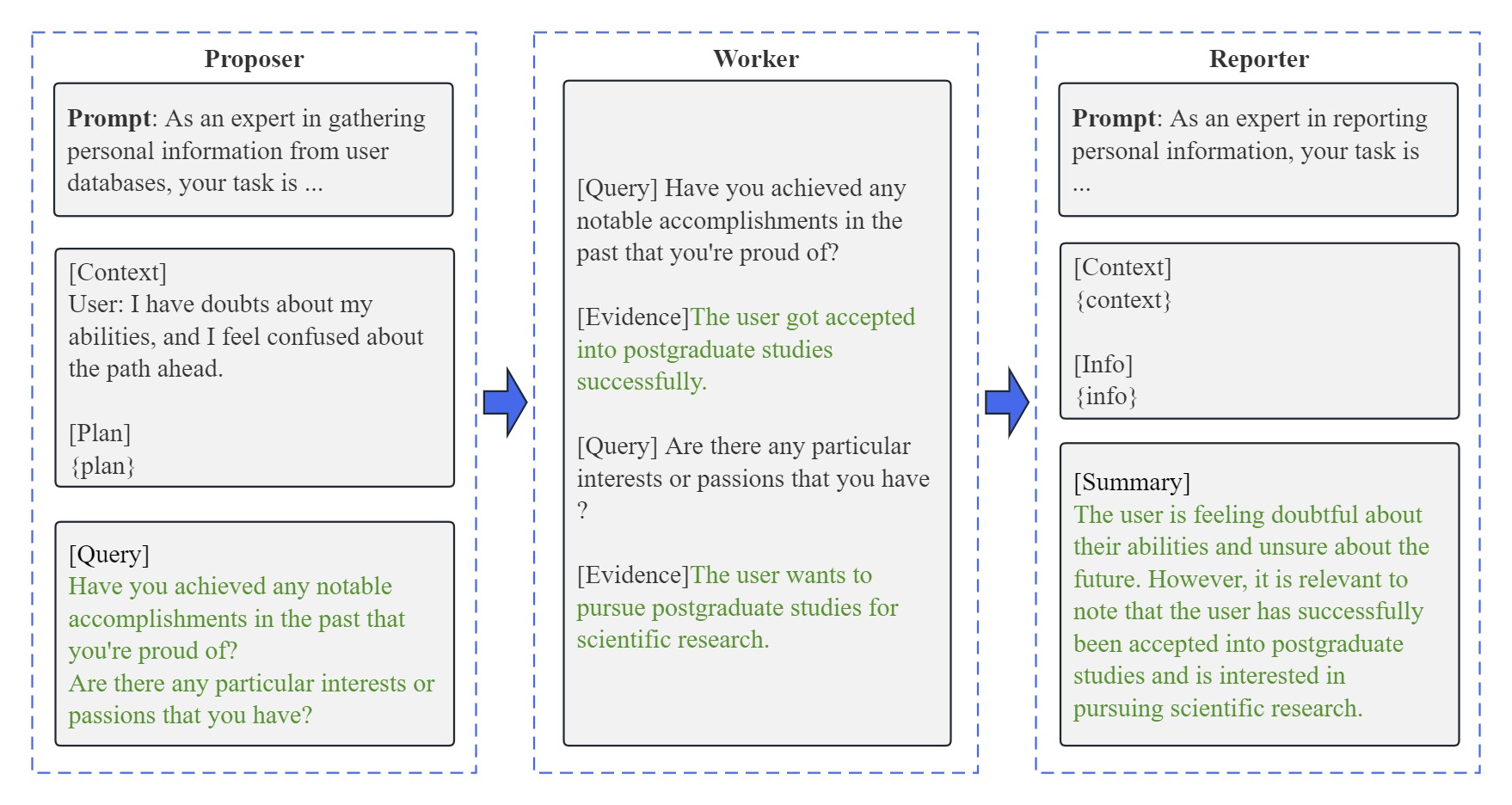}
    \caption{An example of the information retrieval agent.}
    \label{fig:search}
\end{figure}


\paragraph{\textbf{Dialogue Strategy Agent}}
The dialogue strategy agent has two modules: planner and decider. The planner module produces a dialogue strategy plan. The decider module determines the specific strategy action to be taken.

The reasoning process of the planner module includes three steps:
\begin{enumerate}

\item Defining objective: It defines the objective of the dialogue based on the given context, e.g., provide emotional support to the user.

\item Proposing strategy: It proposes a strategy plan based on the defined dialogue objective. The strategy plan can include multiple steps, such as affirming the user's emotions, exploring the causes of the negative emotions, and guiding the user to resolve them.

\item Refining strategy: It refines the strategy plan as the dialogue progresses based on the user's feedback. For example, when the strategy from the previous step is to help the user solve a problem, but the user says, ``I don't want to think about how to solve the problem right now, can you just comfort me?'', the dialogue strategy should be adjusted from problem-solving to providing comfort.

\end{enumerate}

The reasoning process of the decider module also includes three steps:
\begin{enumerate}

\item Analyzing progress: It analyzes the strategy actions taken thus far in the plan. Specifically, it compares the strategies adopted in the conversation with the pre-determined strategy plan to determine which steps of the plan have already been executed.

\item Evaluating outcomes: It analyzes the user's feedback from the conversation to evaluate the effectiveness of the strategies employed; for example, whether the user's emotions have been calmed.

\item Making action decisions: It decides the next strategy action to be taken based on the analysis and evaluation from the previous steps. If the user's emotions haven't been calmed, it continues to address the emotional distress. If the user's emotions have been calmed, it starts guiding the user toward resolving the root problem.

\end{enumerate}


Figure~\ref{fig:strategy} shows an example of dialogue strategy agent. Each module consists of a prompt that describes the task and provides guidance for the reasoning process. The prompt is used as the system prompt for LLM-Base to execute the module's functionality. During conversation, the planner generates a multi-step dialogue strategy plan based on the context of the conversation. Subsequently, the decider determines the specific plan action to take. Finally, the generated text of the strategy action serves as a prompt for guiding LLM-Large to generate a reply that aligns with the specified direction of the strategy.

\paragraph{\textbf{Information Retrieval Agent}}
The information retrieval agent includes three modules: proposer, worker, and reporter. The proposer and reporter utilize prompts for LLM-Base to generate queries and summarize query results, while the worker module executes query operations on the vector database. Figure~\ref{fig:search} provides an example of information retrieval agent.

\textbf{Proposer} is responsible for suggesting which aspects of user information should be retrieved based on the real-time context and strategy plan. Concretely, it includes proposing a list of queries for retrieving historical contexts and user profiles. Each query describes a specific aspect of the user, such as past achievements.

\textbf{Worker} is responsible for executing the query on the vector database and retrieving the corresponding information. During retrieval, OS-1 determines the cosine similarity between the query vector and the vectors of historical context and user profile documents. Once retrieval produces a set of candidate documents, a rank score is calculated for each document and they are sorted to enable selection of the $k$ documents with the highest rank scores. Rank score calculation is similar to that used in a generative agent~\cite{park2023generative}: $S_{\mathit{rank}}=S_{\mathit{similarity}}+S_{\mathit{importance}}+S_{\mathit{recency}}$, where the recency score $S_{\mathit{recency}}$ accounts for the recency of the update time of document creation (the more recent the document, the higher the recency score).

\textbf{Reporter} is responsible for extracting and summarizing relevant information from retrieved documents. Additionally, it creates a description of user information that serves as a prompt for LLM-Large to generate a response.

Finally, we combine the real-time context, the relevant historical contexts, and the user profiles retrieved by the information retrieval agent to produce the personal context. This and the dialogue strategy planned by the dialogue strategy agent are used as prompts for LLM-Large to generate personalized responses (Figure~\ref{fig:response_generation}).

\subsection{Implementation} \label{subsec:eyewear-system}
\subsubsection{Hardware Design}

\begin{figure}[htbp]
    \subfigure[Diagram of the left eyewear leg] 
    {
        \begin{minipage}{0.3\textwidth}
            \centering        
            \includegraphics[width=1\textwidth]{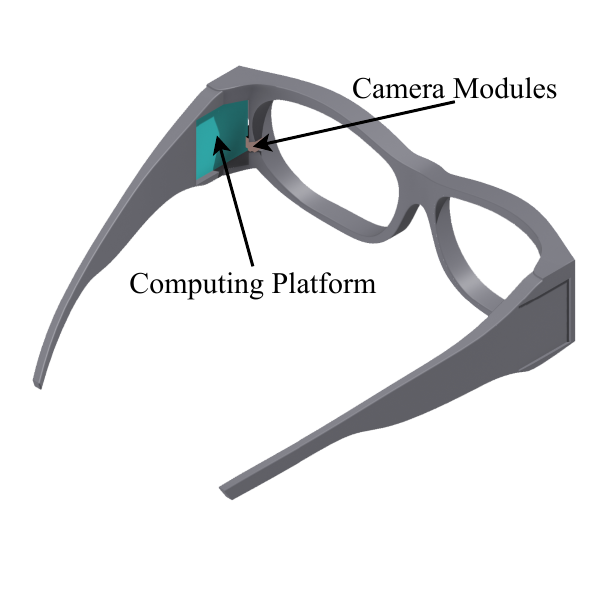}
        \end{minipage}
    }
    \subfigure[Diagram of the right eyewear leg] 
    {
        \begin{minipage}{0.3\textwidth}
            \centering        
            \includegraphics[width=1\textwidth]{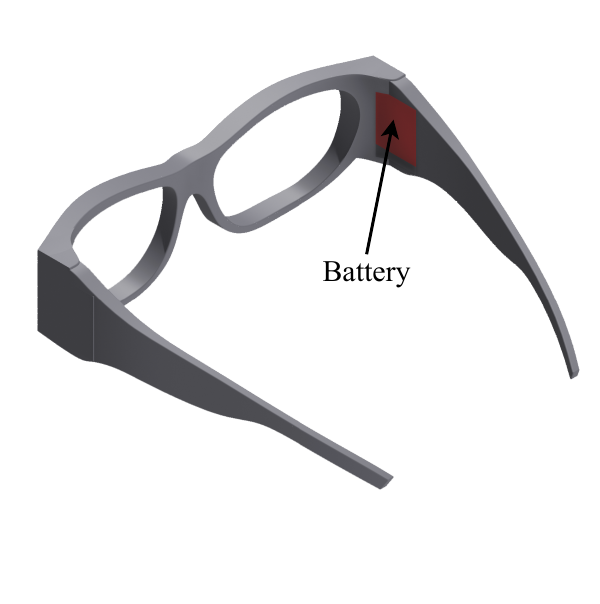}
        \end{minipage}
    }
    \subfigure[Actual Wearing Photo]
    {
        \begin{minipage}{0.3\textwidth}
            \centering    
            \includegraphics[width=1\textwidth]{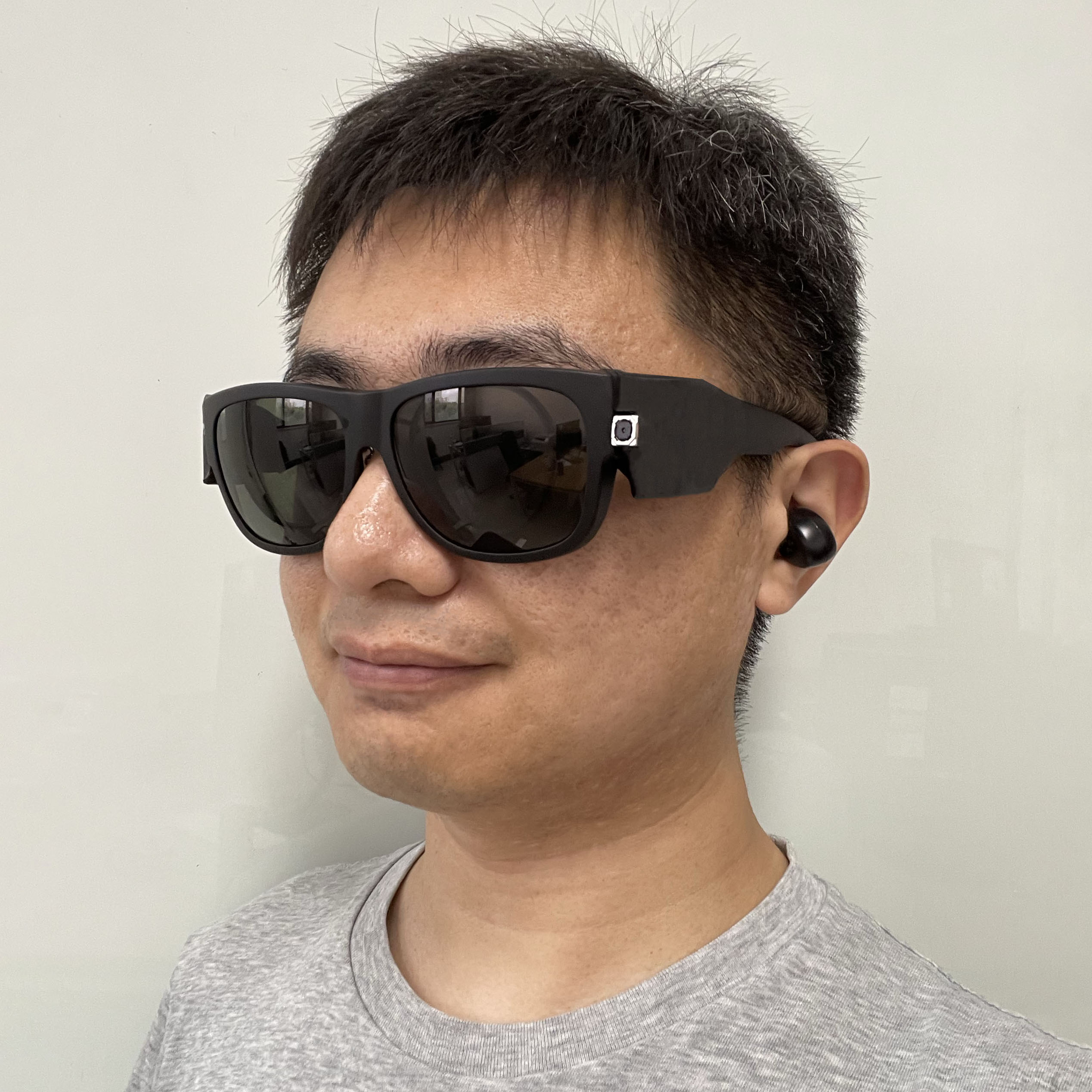}
        \end{minipage}
    }
    \caption{Prototype diagram and wearing photo of the eyewear.}
    \label{fig::eyewear}
\end{figure}

In the smart glasses designed for OS-1, computing components, and a power source are seamlessly into the glasses frame.
Figure~\ref{fig::eyewear} illustrates the eyewear hardware prototype that brings our vision to life. 

The Snapdragon Wear 4100+ \cite{qualcommSnapdragon} is used as the computing platform and is directly integrated into the left arm of the glasses. This platform's  processing speed is adequate for real-time data processing and execution of sophisticated algorithms, such as eye tracking and scene capturing.

The eyewear hardware is equipped with two cameras: an 8-megapixel (MP) scene camera and a 5-megapixel (MP) eye camera. 
The scene camera captures the surrounding scene images, providing visual context to the system.
The eye camera records eye videos, supporting eye tracking. 

The eyewear frame design houses the eye and scene cameras within its left arm. The cameras are unobtrusive and aligned with the user's field of view.
To provide a well-balanced and comfortable fit, we integrated the battery into the right arm of the glasses, thereby balancing the frame.

In addition to the body of the eyewear, we also incorporate a monaural Bluetooth earphone that is used to record audio of the user and environment. A speaker is used to produce verbal responses.

\subsubsection{Software Design}

In this section, we elaborate on the software design aspects of the eyewear system. The system operates on Android 8.1, providing a platform for communication between the user and the cloud services. 
Initially, the user is required to configure the WiFi connection to access the cloud and enable uninterrupted communication.
The software has four functions: capturing audio, scene images, eye orientations, and playing the audio output of responses received from the cloud server.
\begin{itemize}

    \item \textbf{Audio:} The eyewear system continuously captures audio from the user's surroundings, which is streamed to the cloud in real-time. In the cloud, a voice recognition system processes the audio stream, converting it into text. 

    \item \textbf{Image:} The eyewear system periodically captures 640$\times$480 scene images at specific time intervals (every 10 seconds in this work). To optimize data transmission, the captured images undergo JPEG compression before being uploaded to the cloud. Once uploaded, the cloud performs feature extraction on the images, allowing insight into the user's current environment.

    \item \textbf{Eye-tracking:} An eye-tracking algorithm similar to Pupil Invisible \cite{tonsen2020highlevel} is run on the eyewear system. The algorithm can provide the position of the user's gaze on scene images.

    \item \textbf{Playback:} The eyewear system plays the human-like audio response generated from the cloud.

\end{itemize}

\subsubsection{Cloud Services}


The cloud services consist of five components, each capable of handling multiple processes concurrently to support simultaneous interactions with multiple users. Redis~\cite{redisRedis} queues are used for communication among these services.
\begin{itemize}

    \item \textbf{Data Server:} The data server is responsible for facilitating communication with the eyewear. It is built on the FastAPI framework~\cite{tiangoloFastAPI} and has two key interfaces. The first allows uploading data, including timestamps, audio, images, and other relevant information. Upon receipt, these data are placed in the appropriate queue, awaiting processing. The second interface returns generated audio replies. It retrieves audio from the response queue and streams it to the user's eyewear through the Starlette framework \cite{starletteStarlette}.

    \item \textbf{Image Server:} The image server component retrieves images from the queue and processes them using the LLaVA~\cite{liu2023llava} model for content recognition. Specifically, the LLaVA-7B-v0 model is employed, with parameter settings as follows: max\_new\_tokens = 512 and temperature = 0.

    \item \textbf{Audio Server:} For each online user, a dedicated thread is created to handle the audio input. This thread continuously receives audio data from the users' eyewear system and uses Whisper~\cite{radford2023robust} for speech recognition.

    \item \textbf{Chatbot Server:} The chatbot server serves as the core service within the cloud, generating responses based on the user's surrounding environment and conversation content. The responses include textual content, as described in Section~\ref{subsec:cloud-workflow}. 


    \item \textbf{TTS Server:} The TTS server converts textual responses into the audio format. This component uses a commercial text-to-speech service \cite{alibabacloudIntelligentSpeech} for efficient and high-quality audio synthesis.

\end{itemize}

The processing time for the cloud services is approximately 1.82 seconds, which is at the most common pause time in human conversation (1-3 seconds)~\cite{elsner2010disentangling}, allowing for natural communication with OS-1.



\section{Evaluation}

\label{sec:evaluation}
This section evaluates the performance of OS-1 empowered by effective personal context capturing, which is designed to cater to diverse users with varying profiles who engage in various conversation scenarios during their daily lives. To this end, we first consider a variety of conversation situations and simulated users with various profiles in a controlled laboratory setting. Then, we recruit volunteers to participate in pilot studies for approximately 14 days to examine the long-term effectiveness when OS-1 is used in real-world scenarios.

\subsection{In-lab Experiments}
For the in-lab experiments, we first outline the experimental settings to simulate various daily-life scenarios and users with diverse social backgrounds and personalities. Then, we compare the performance of our proposed system, OS-1, with that of the baseline methods without considering personal context. Lastly, we use a case study to further explain why OS-1 outperforms the baseline methods.

\subsubsection{Experimental Setup}
\paragraph{(1) User Simulation} To verify OS-1's ability to adapt to diverse users, we use GPT-4 to simulate virtual users with varying personalities, social backgrounds, and experiences, which follows the approach~\cite{aher2023using}. In particular, we create 20 distinct virtual users consisting of 10 males and 10 females, ranging in age from 15 to 60. Each virtual user is assigned a name randomly selected from the U.S. 2010 Census Data~\cite{bureau2010name}. Also, we assign each user a personality based on the Myers-Briggs Type Indicator (MBTI)~\cite{myers1962mbti}. To make the virtual users more realistic, we provide each virtual user with an occupation, preferences, and habits, along with daily routines tailored to their individual characteristics.

\paragraph{(2) Visual Scene Simulation}
We use GPT-3.5 to directly simulate the 20 user's daily visual scenes at a given moment. The visual scenes represent the visual surroundings perceived by users, and they are represented as a four-tuple, including time, location, activity, and a brief text description of what the user perceives. For example, a college student, Benally majoring in Chemistry, might experience a visual scene of <2023-10-02 Monday 9:00-12:00, Chemistry Lab, Attending lectures and practicals, ``A table filled with beakers and test tubes.'' >. 

In total, we simulate a total of 80 daily visual scenes for each user, with 8 scenes per day and a duration of 10 days.


\paragraph{(3) Dialogue Simulation}
We randomly select three daily visual scenes for each user and ask the user to initiate a conversation with OS-1 based on the visual scene. Each conversation consists of three rounds. This way, we get each user's personal context, consisting of the simulated speech and their daily visual surroundings. Then, we cluster the personal context and summarize the historical context with a few sentences to describe it. Furthermore, we distill the user profile using the historical context.

\paragraph{(4) Test Scenario Simulation}
We also create the test scenarios to verify OS-1's capability to reach better grounding by utilizing their context. To achieve this, we recruit a human experimenter to review the virtual users' personal context and instruct the experimenter to specify a chat topic and a brief text that describes a visual scene. For example, a chat topic may be ``dinner recommendations'' and a visual scene may be ``a commercial street with a pizza stand''.

\paragraph{(5) Evaluation Measures}
There are no benchmark measures that we could adopt to evaluate OS-1 directly. 
As indicated by previous studies~\cite {1905.04071,adiwardana2020humanlike,hashimoto2019unifying}, evaluating dialogue systems is a challenging task. Moreover, few previous studies considered personal context-empowered dialogue systems integrated with smart eyewear. Following the idea of using the key elements of human likeness as criteria for assessing the quality of conversational responses~\cite{adiwardana2020humanlike}, this work also proposes customized human evaluation metrics to evaluate the performance of OS-1. Specifically, we propose to use \textit{Grounding} score as the first metric to assess the overall quality of OS-1's response content. The \textit{Grounding} score indicates how well OS-1 can establish common ground with its users, thus generating human-like responses that meet users' expectations given the personal context. To provide a more detailed description, we break down the \textit{Grounding} score into the following three dimensions to assess the quality of response content and the long-term interactive effect.
\begin{itemize} 
      \item \textit{Relevance:} Following~\cite{liang2021towards}, the relevance score is used to test the correlation between the response and the user's speech and their in-situ environment, including the location, visual surroundings, current activity, and time. 
  
      \item \textit{Personalization:} The personalization score determines how closely the response relates to the user's specific information, including their profile and the semantics derived from what they are currently viewing and chatting about, as well as their past interactions with OS-1.  
  
      \item \textit{Engagement:} The engagement score measures how interested a user is in the response and whether the response will lead to further conversation. This is similar to previous studies~\cite{li2019acute, liang2021towards}.

\end{itemize}
These three metrics are supplementary to \textit{Grounding} score, and ideally, the higher score in all three metrics should result in a higher \textit{Grounding} score.

This work adopts the widely-used 5-point Likert scale~\cite{revilla2014choosing} to evaluate the responses from OS-1 and the baseline methods. Also, to mitigate the possible bias from human raters, we involve 15 human raters and ensure that each response is evaluated by at least three of them. We then adopt the mean value of the ratings.

\paragraph{(6) Baseline Methods} 
As there are no previous methods that can be directly compared to OS-1, we conduct ablation studies to evaluate its performance. The ablation studies have two purposes. First, they evaluate the OS-1's ability to establish common ground with users by incorporating their personal context and generate more personalized responses. Second, they help us quantify the contribution of real-time, historical, and user profile context to establishing common ground.
\begin{itemize}

\item \textit{w/o} P: This method solely relies on the real-time and historical context to boost context-aware dialogue generation. The user profile is omitted.
			
\item \textit{w/o} PH: This method only leverages the real-time context to enhance the context-aware dialogue generation. It omits historical context and user profiles.
			
\item \textit{w/o} PHR: This method uses an LLM to produce responses during interaction with users, omitting any personal context.
			
\end{itemize}

\subsubsection{Overall Performance}

\begin{figure}[htbp]
    \centering
    \begin{minipage}[t]{0.48\textwidth}
        \centering
        \includegraphics[width=1\textwidth]{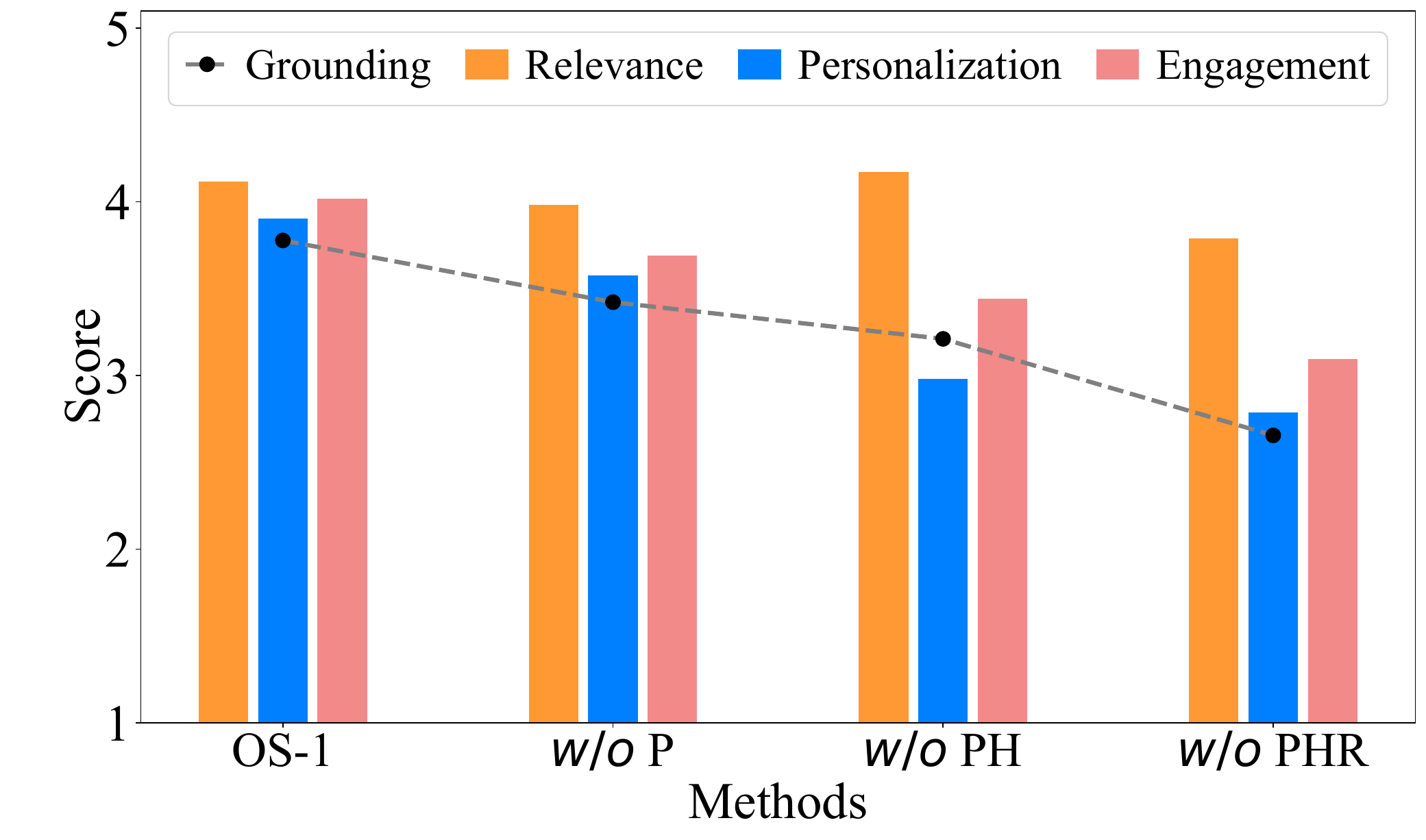}
        \caption{Performance comparison of OS-1 with the baseline methods}
        \label{fig::inlab-overall}
    \end{minipage}
    \hspace{0.02\textwidth}
    \begin{minipage}[t]{0.48\textwidth}
        \centering
        \includegraphics[width=1\textwidth]{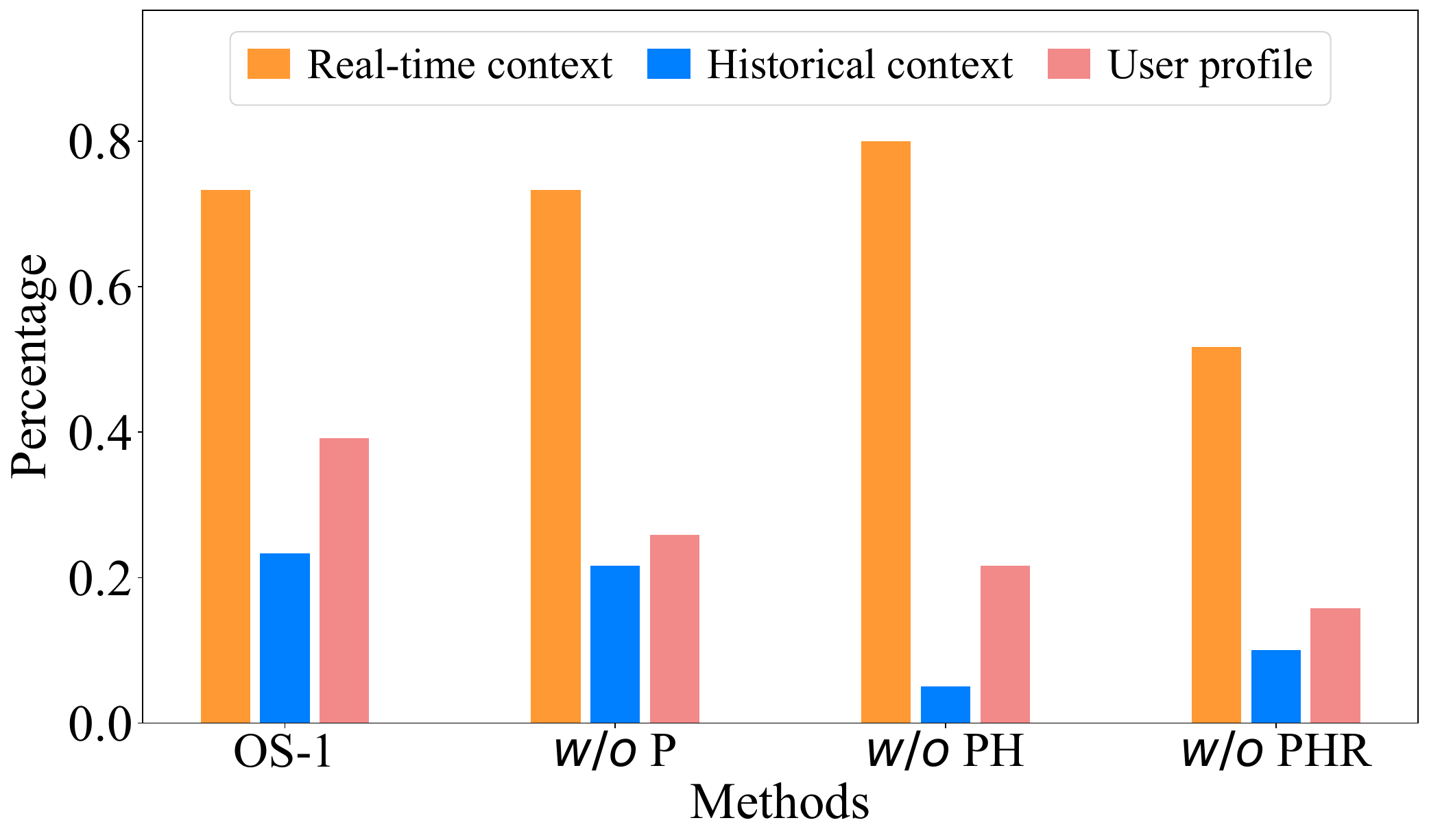}
        \caption{Percentage of various factors that contribute to better grounding}
        \label{fig::inlab-context}
    \end{minipage}
    
\end{figure}

Figure~\ref{fig::inlab-overall} shows the performance of different methods in terms of \textit{Grounding}, \textit{Relevance}, \textit{Personalization}, and \textit{Engagement} score based on the human raters. As we can see, OS-1 achieves the highest scores among the four methods. Compared with the \textit{w/o} PHR, OS-1 improves the \textit{Grounding} score by 42.26\%. Also, OS-1 substantially improves the performance by 8.63\%, 40.00\%, and 29.81\% in \textit{Relevance}, \textit{Personalization}, and \textit{Engagement}, respectively. 

Next, we further investigate the factors that aid in better grounding from the viewpoint of human raters. We ask the human raters to review all the responses generated by various methods and identify the factors that contribute to good grounding for each response. The raters consider three aspects: the proposed real-time context, historical context, and user profile. Also, the raters are allowed to select multiple factors that lead to good grounding. We then calculate the percentage of the number of each factor selected by the raters out of all the selected responses. The results are presented in Figure~\ref{fig::inlab-context}. We observe that the personal context plays a significant role in building good grounding. Specifically, we find that (1) the percentage of the methods that include the real-time context is higher (0.73, 0.73, 0.80 for OS-1, \textit{w/o} P, and \textit{w/o} PH, respectively) compared to those without personal context (0.51 for \textit{w/o} PHR). (2) similarly, methods that include historical context have a higher percentage (0.23 and 0.21 for methods OS-1 and \textit{w/o} P, respectively) than those without such context (0.05 and 0.10 for methods \textit{w/o} PH and \textit{w/o} PHR, respectively). and (3) the percentage of methods that include the user profile is higher compared to those without this kind of context (0.39 for OS-1, compared to 0.26, 0.22, 0.16 for \textit{w/o} P, \textit{w/o} PH, and \textit{w/o} PHR, respectively). 

\subsubsection{A Case Study}
\begin{figure}[htbp]
      \centering
      \includegraphics[width=1\textwidth]{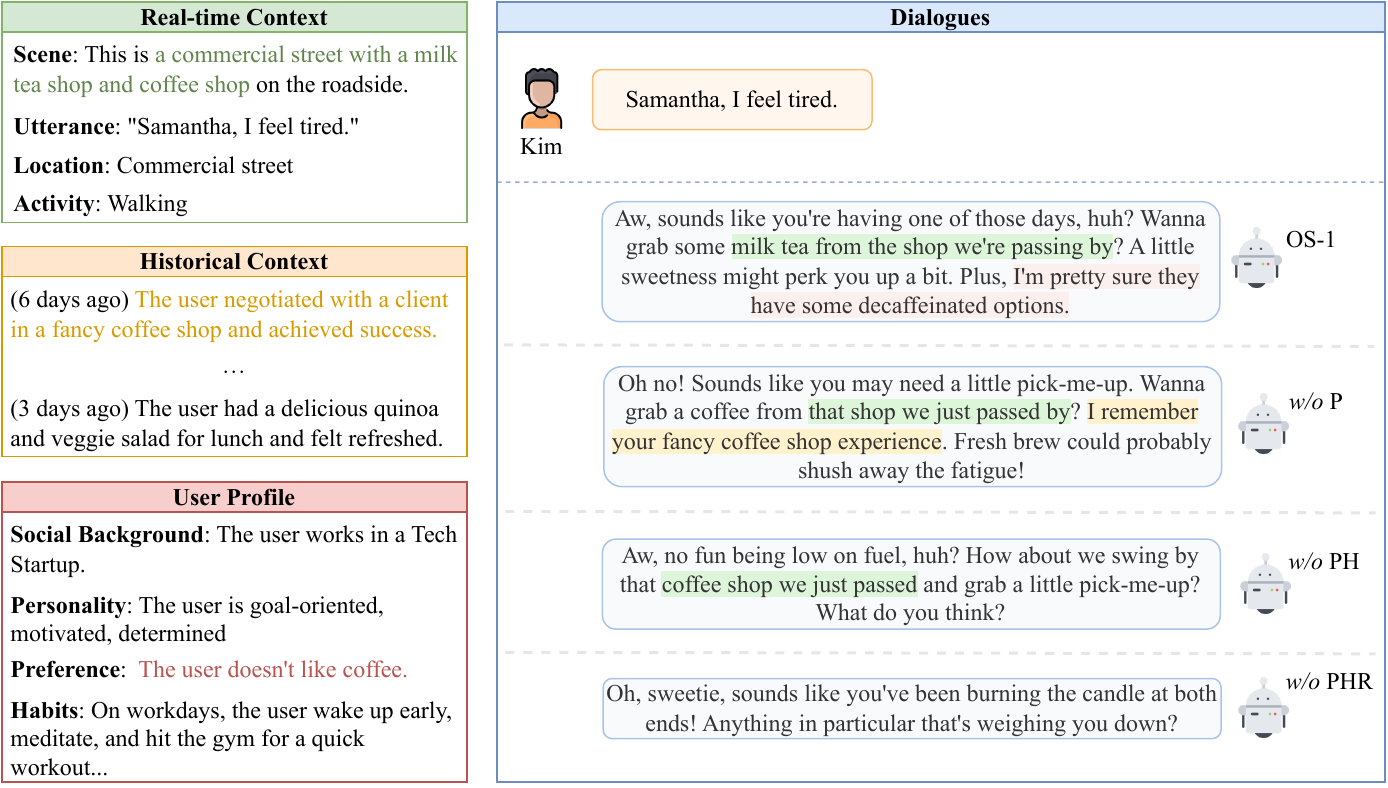}
      \caption{A case of shop recommendation. Left: personal context captured by OS-1. Right: four dialogue sessions between a simulated user and four systems}
      \label{fig::inlab_case1}
\end{figure}

We provide a case study to offer further insights regarding why OS-1 outperforms the baselines for personalized dialogue. Figure~\ref{fig::inlab_case1} shows the dialogue sessions between a simulated user named Kim and four systems, including the proposed OS-1 and three baseline methods. As shown in the real-time context, Kim is walking along a commercial street with a coffee shop and a milk tea shop, an important piece of real-time context. Historical context reveals that Kim has been to a coffee shop for business recently. User profiles reveal that Kim dislikes coffee. Compared with the three baseline methods, we observe that OS-1 provides the most appropriate response by using both real-time context relevant information (highlighted in green) and user profile context-relevant information (highlighted in red). In contrast, other baseline methods miss one or more, such as \textit{w/o} P, which retrieves historically relevant information accurately but misses user profiles, and \textit{w/o} PHR lacks the three pieces of context information.

\subsection{Pilot Study}
In addition to laboratory studies, we also performed a two-week pilot field study to observe the behavior of OS-1 in the real world. 
First, we determine whether OS-1 is capable of extracting the profiles and long-term historical contexts of users through multiple interactions. We then assess the OS-1's ability to establish common ground with its users. In particular, \textit{Grouding}, \textit{Relevance}, \textit{Personalization}, and \textit{Engagement} scores were measured. Second, we describe two potential downstream applications for which OS-1 would be appropriate: providing emotional support and personal assistance. All experimental procedures are approved by the ethics committee at our university.

\subsubsection{Procedure of Pilot Study}
We recruited volunteers from our university to participate in the pilot study. Prior to the pilot study, we inform the participants that the glasses will perceive their daily visual scenes and audio, and the researchers will examine their daily chat logs recorded in the eyewear system only with their permission. The raw sensed image and audio data will be removed right after feature extraction, and only anonymized semantics are transmitted and stored securely in the cloud. All participants were aware of this procedure and signed consent forms prior to their experiments.
We also provide each participant with instructions on how to use the OS-1, including starting a conversation, turning off the system, and reviewing the conversation history using the designed web service.

The pilot study consists of two phases, and each lasts 7 days, with slightly different purposes. In the first phase, we recruit 10 volunteers (aged 22-28, 6 males and 4 females, referred to as P1 to P10 in the following text) plus 3 authors to attend the pilot study. The main reason for involving three authors is that they can collect first-hand user experience and make necessary and timely adjustments to the system pipeline. Those 3 authors only show up in the first-phase studies and are excluded from the second phase. As we still face the problem of the limited concurrency ability of the system, we reserve varying time slots for different participants. 
After completing the first phase, we spend one month improving the system concurrency as well as the hardware usability. Then, we move on to the second-phase pilot study with 10 participants aged 22-29, 7 males and 3 females, referred to as P11 to P20 in the following text. In the second phase, the participants can use the system anywhere and at any time. 

After completing the daily experiments in both phases, we ask the participants to review the responses generated by OS-1 and score them using the same criteria as in the laboratory experiments, i.e., \textit{Grouding}, \textit{Relevance}, \textit{Personalization}, and \textit{Engagement} score. We also make a slight adjustment to make the score more suitable for in-field settings. Instead of using the 5-point Likert scale used in laboratory settings, we expand the evaluation scale to an 11-point Likert scale~\cite{lewis2017user, leung2011comparison}. This allows us to collect more fine-grained scores, enabling us to track the long-term score changes when OS-1 is used in the real world.

In both phases, we ask the participants to use the system for at least 30 minutes per day and encourage them to use it as long as possible. In the first phase, we collect 26.85 minutes of conversation per day on average, comprising 53.70 utterances from both the participants and OS-1. In contrast, in the second phase, we collect 27.64 minutes of conversation per day on average, comprising 65.62 utterances, which is higher than that of the first phase. This is in line with our expectations, as we improve the system stability and concurrency significantly after the first phase, making participants' interactions with OS-1 smoother, resulting in more conversations between the participants and OS-1.

\subsubsection{Performance of OS-1 in Pilot Study}
\begin{figure}[htbp]
\centering
\begin{minipage}[t]{0.48\textwidth}
\centering
\includegraphics[width=1\textwidth]{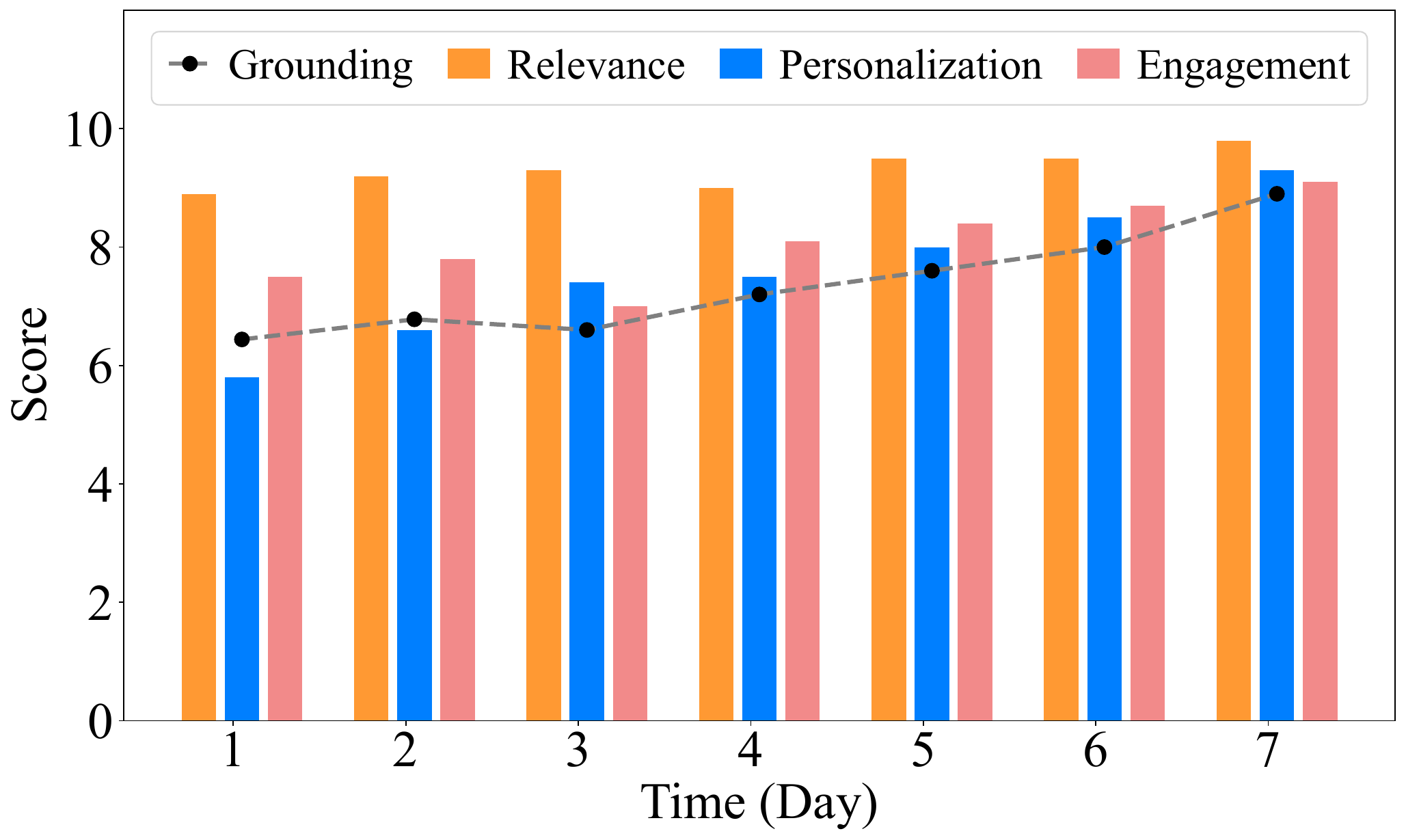}
\caption{The average evaluation scores of all participants in phase 1.}
\label{fig::human_eval_p1}
\end{minipage}
\hspace{0.02\textwidth}
\begin{minipage}[t]{0.48\textwidth}
\centering
\includegraphics[width=1\textwidth]{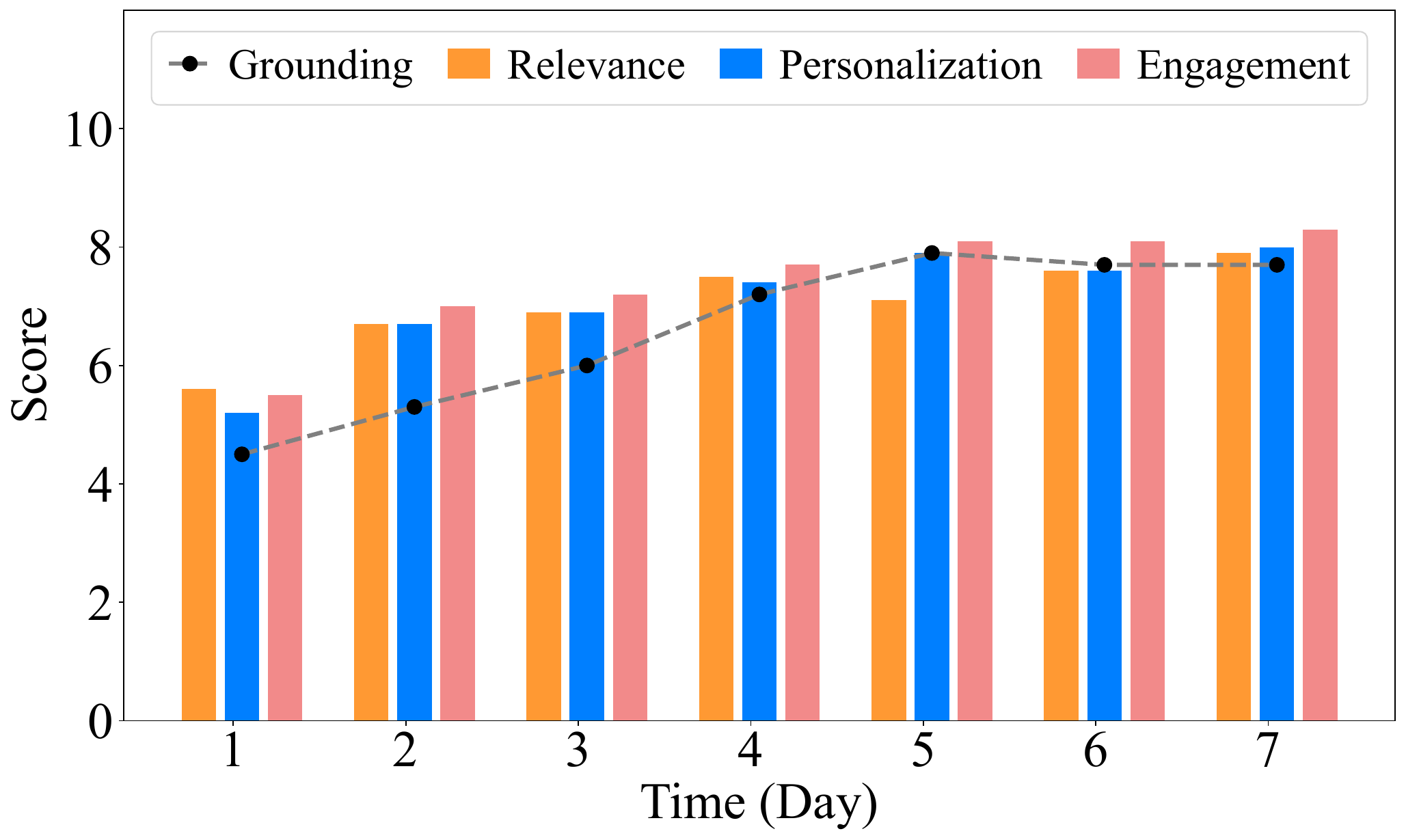}
\caption{The average evaluation scores of all participants in phase 2.}
\label{fig::human_eval_p2}
\end{minipage}
\end{figure}

Figure~\ref{fig::human_eval_p1} and Figure~\ref{fig::human_eval_p2} depict the average evaluation scores of the 10 participants over the 7-day period in the first and second phases, respectively. Clearly, the participants find OS-1's responses to be relevant, personalized, and engaging, with most scores higher than 5. Moreover, despite the small fluctuations, all scores show a consistently increasing pattern over the 7 days.
This indicates that OS-1 is able to generate responses tailored to each participant's personality throughout time. More importantly, as we can see, the \textit{Grounding} score also shows an increasing trend over time throughout the two pilot phases, which indicates that OS-1 is capable of continuously reaching common ground with users through long-term interactions. As a result, users perceive that OS-1 understands them better over time by making conversations more relevant, personalized, and engaging.

\begin{figure}[htbp]
\centering
\begin{minipage}[t]{0.48\textwidth}
\centering
\includegraphics[width=1\textwidth]{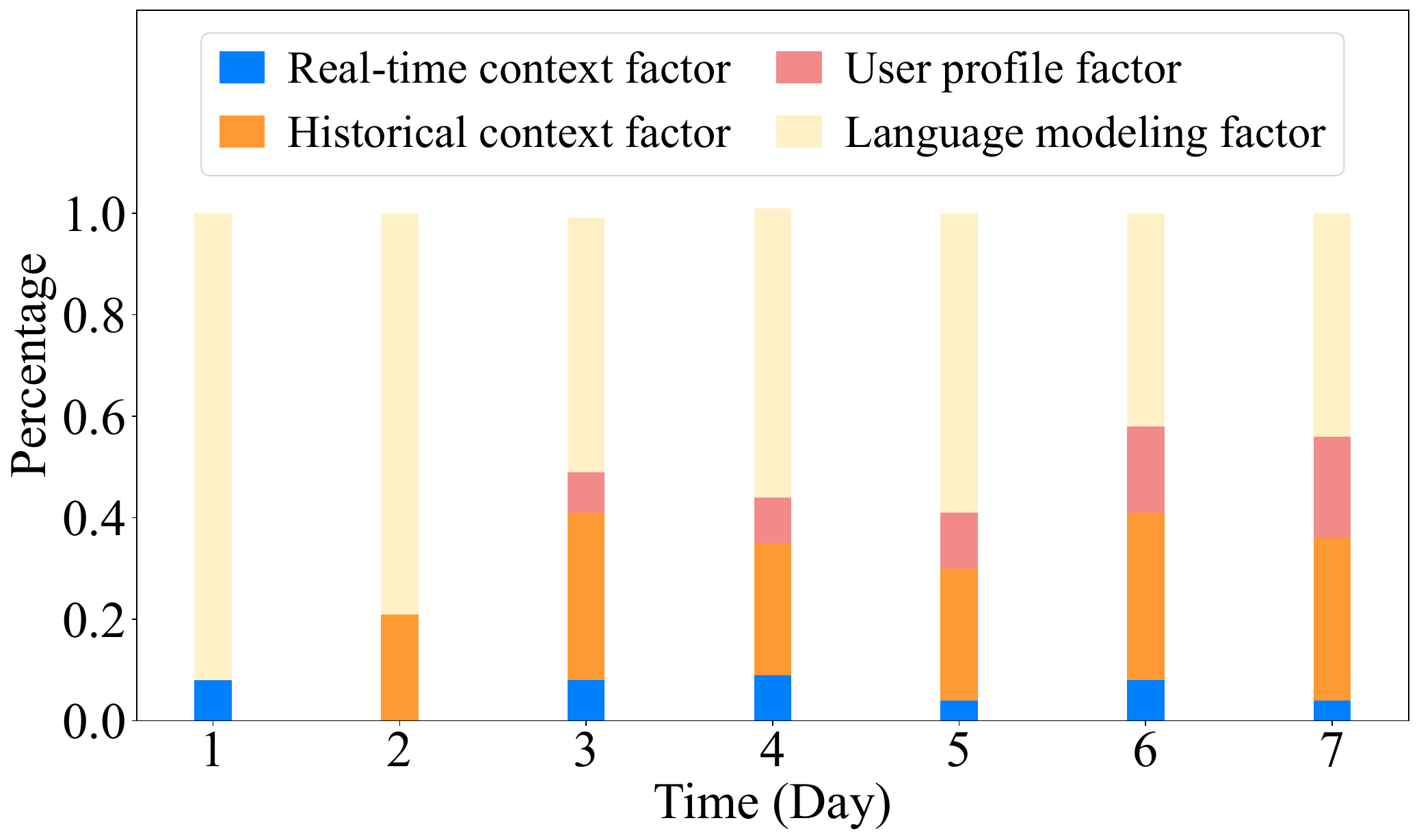}
\caption{The daily percentages of the four factors in phase 1.}
\label{fig::success_factors_p1}
\end{minipage}
\hspace{0.02\textwidth}
\begin{minipage}[t]{0.48\textwidth}
\centering
\includegraphics[width=1\textwidth]{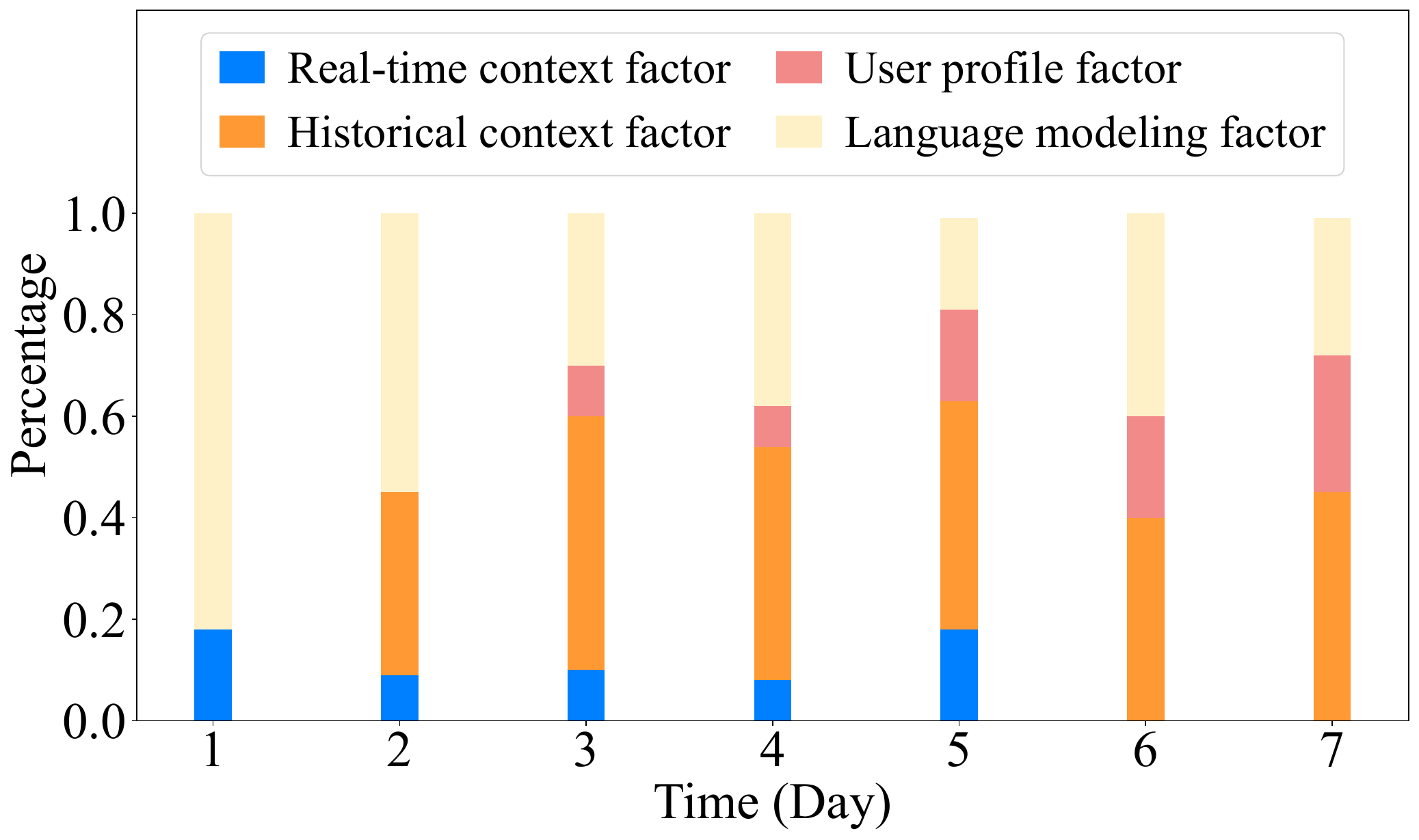}
\caption{The daily percentages of the four factors in phase 2.}
\label{fig::success_factors_p2}
\end{minipage}
\end{figure}

To evaluate whether personal context contributes to a better common ground with OS-1 and leads to more personalized responses, we ask the participants to pick the daily response that best reflects OS-1's understanding of them. They are also asked to specify the reasons for their choices. We provide four options with three personal context-related factors and one LLM-related factor.
\begin{itemize}

\item \textbf{Real-time context factor} The response is linked to the scene and the conversation the user had at a specific time. 

\item \textbf{Historical context factor}: The response is retrieved from the historical semantics stored in the database.


\item \textbf{User profile factor}: The response is closely related to the summarized user profile, such as personality and habits. 

\item \textbf{Language modeling factor}: The response is generated solely by the LLM without taking into account personal context.

\end{itemize}
A human examiner reviews the selected responses and the corresponding reasons, and manually chooses to assign one of the above options to each response that can best explain why the participants select it. We calculate the percentage of the number of each factor selected out of the number of all the selected responses. The results are provided in Figures~\ref{fig::success_factors_p1} and \ref{fig::success_factors_p2}. The two figures show the daily percentage contribution of each factor to establishing common ground during the 7-day period in the two pilot phases. Ideally, a higher percentage of a factor leads to a higher contribution to a user-preferred response. 
We observe that the percentages of the personal context-related factors increase over time, e.g., the historical content factor and the user profile factor, while that of the LLM factor decreases. This also suggests that OS-1 can utilize the user historical contexts and learn user profiles from past interactions, and generate more personalized responses.

Next, we present three concrete cases to explore how personal context-related factors contribute to personalized dialogue responses. 

\begin{figure}[htbp]
      \centering
      \includegraphics[width=1\textwidth]{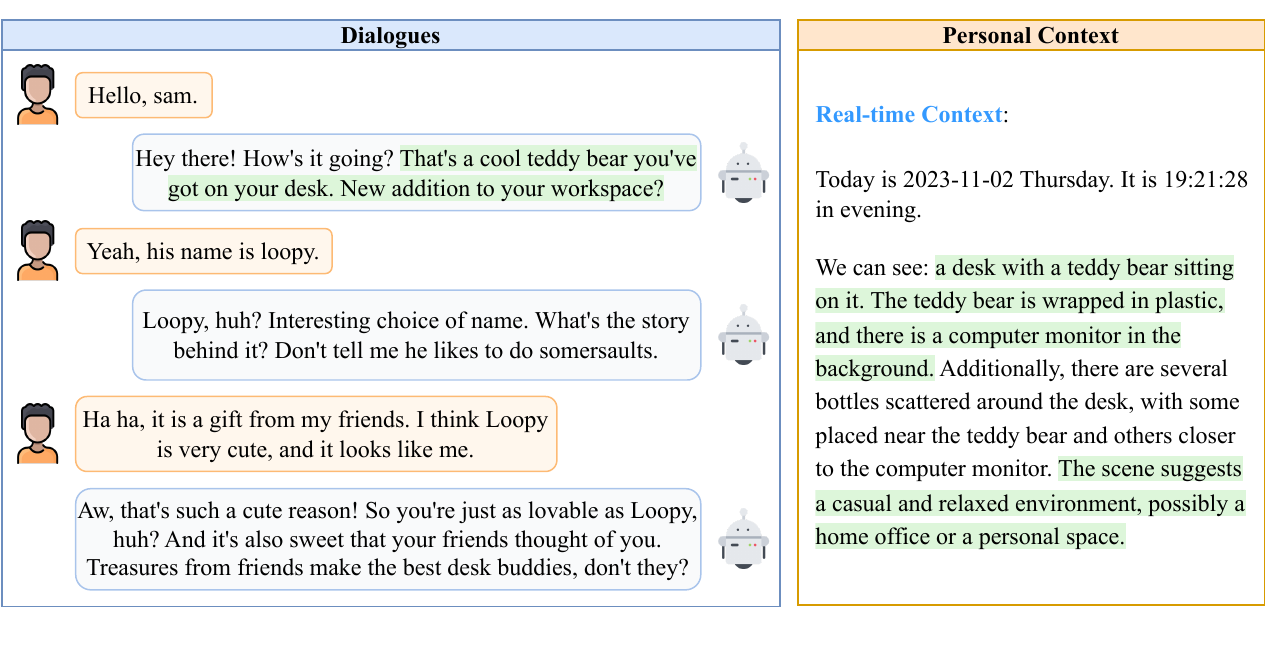}
      \caption{A case that OS-1 is real-time context-aware (Participant P11, Day 1). Green highlights indicate the real-time context-relevant information. }
      \label{fig::episodic_case}
  \end{figure}

\paragraph{(1) A Real-time Context-Aware Case}
Figure~\ref{fig::episodic_case} shows a case when the real-time context factor plays a significant role in the dialogue. Specifically, OS-1 observes that Participant P11 places a Teddy bear on their desk, and thus her greetings involve the information related to that particular visual scene, i.e., \emph{a cool teddy bear}.

\begin{figure}[htbp]
      \centering
      \includegraphics[width=1\textwidth]{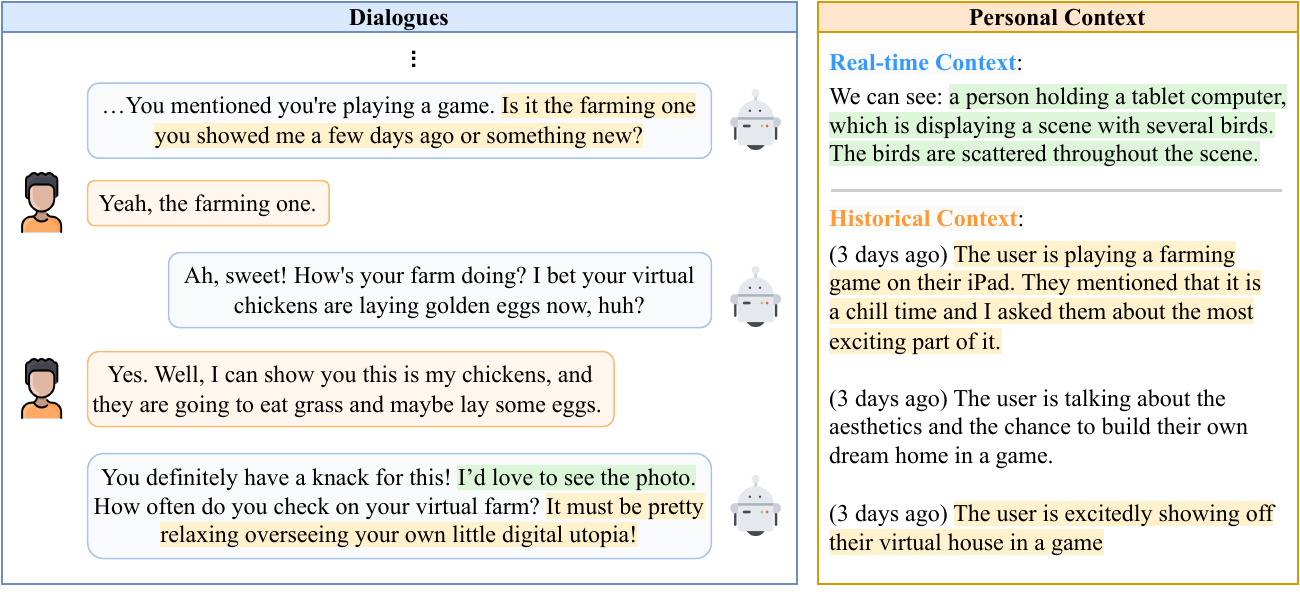}
      \caption{A case that OS-1 is historical context-aware (Participant P16, Day 4). Yellow highlights indicate the historical context relevant information.}
      \label{fig::historical_case1}
\end{figure}

\paragraph{(2) A Historical Context-Aware Case}
Historical context can ensure that the conversations are coherent and consistent over time. We present one example in Figure~\ref{fig::historical_case1}. On day 4, Participant P16 tells OS-1 about playing a game. OS-1 immediately guesses that P16 might be playing the farming game the participant played a few days ago.  They then engage in a coherent conversation about the game. OS-1 also recalls from a previous conversation that P16 has described playing this farming game as a `chill time'. In response, OS-1 comments, `It must be pretty relaxing overseeing your own little digital utopia.'

\begin{figure}[htbp]
      \centering
      \includegraphics[width=1\textwidth]{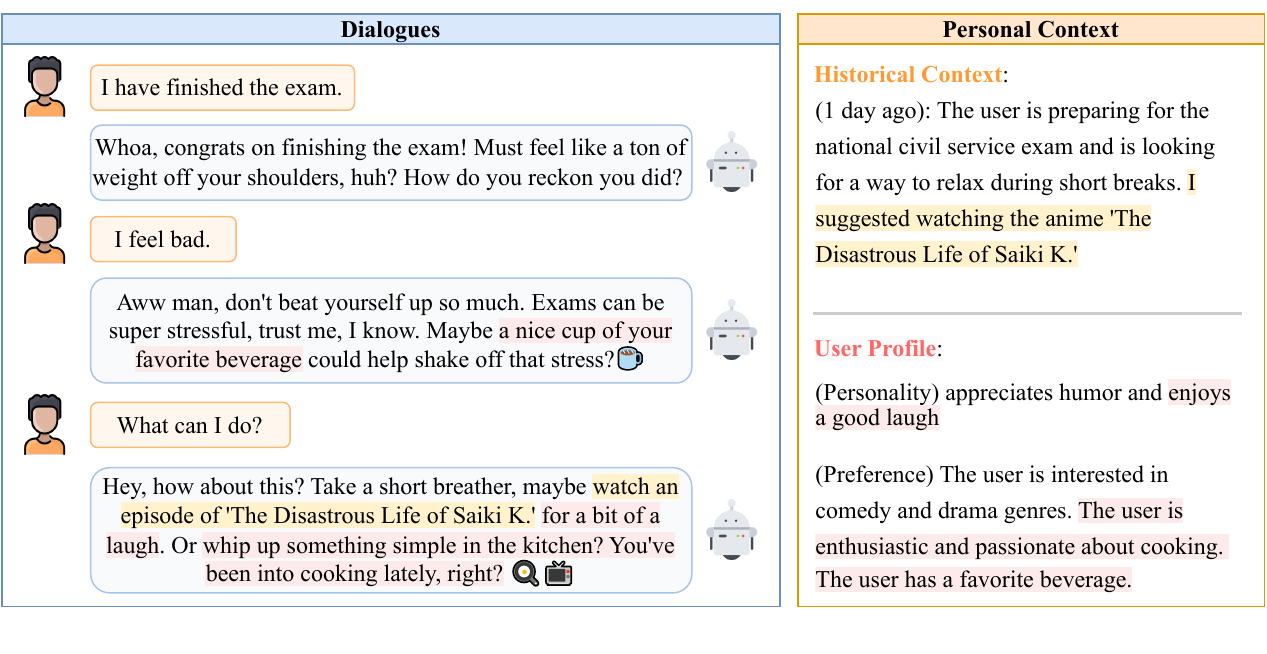}
      \caption{A case that OS-1 is user profile context-aware (Participant P12, Day 4). Yellow and red highlights indicate the historical context and user profile relevant information, respectively. We use the same colouring scheme to label the three types of information hereinafter.}
      \label{fig::profile_case1}
\end{figure}

\paragraph{(3) An User Profile Context-Aware Case}
The user profile allows OS-1 to understand the user's personality traits, social background, preferences, and habits. This information helps OS-1 to create user-specific responses. An example of this can be seen in Figure~\ref{fig::profile_case1}, where OS-1 provided emotional comfort to P12. In this example, OS-1 learned from the historical context that `the exam' mentioned by P12 referred to the national civil service exam that he had been preparing for recently. When P12 expressed feeling bad, OS-1 used the user profile to learn that P12 had a favorite beverage and was passionate about cooking. Based on the historical context, OS-1 also knew that it had recently recommended an anime to P12. Therefore, OS-1 suggested ways to relieve P12's stress based on this information.

\subsubsection{Applications}

\paragraph{(1) Emotional Support}
Research in sociology and psychology has revealed that human emotions have a significant impact on various aspects of our daily lives. Emotions can affect our thoughts and behaviors~\cite{costa2016emotioncheck}, decision-making~\cite{ruensuk2020you}, and physical and mental health~\cite{di2018emotion, tugade2004resilient}. With this in mind, we believe that OS-1 has the potential to provide emotional support for users. As a personal context-powered system, OS-1 shows the ability to understand and connect with users on a deeper level. Through our user interviews, we find that 8 out of 10 participants believe that OS-1 can provide valuable emotional support. 

\begin{figure}[htbp]
      \centering
      \includegraphics[width=1\textwidth]{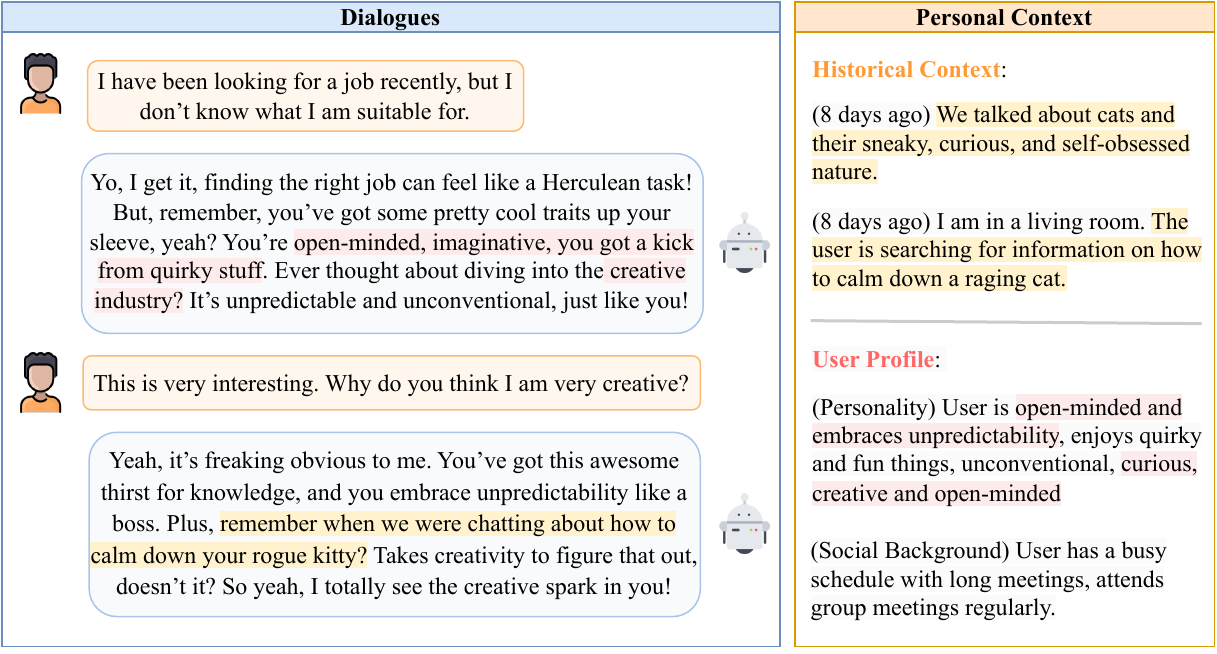}
      \caption{A case when OS-1 assists the participant to escape from negative emotions (Participant P5, Day 7).}
      \label{emotion_case1}
\end{figure}

Figure~\ref{emotion_case1} illustrates a situation where Participant P5 shares anxiety about job hunting with OS-1. Using the user profile built from their past interactions, OS-1 encourages P5 to act as an open-minded, imaginative, and creative person. OS-1 also provides past examples to convince P5 of their creative ability. Through our daily surveys, P5 reports satisfaction about the emotional support provided by OS-1, as P5 believes that OS-1 can demonstrate its creativity by citing past events, which makes P5 more convinced.

\begin{figure}[htbp]
      \centering
      \includegraphics[width=1\textwidth]{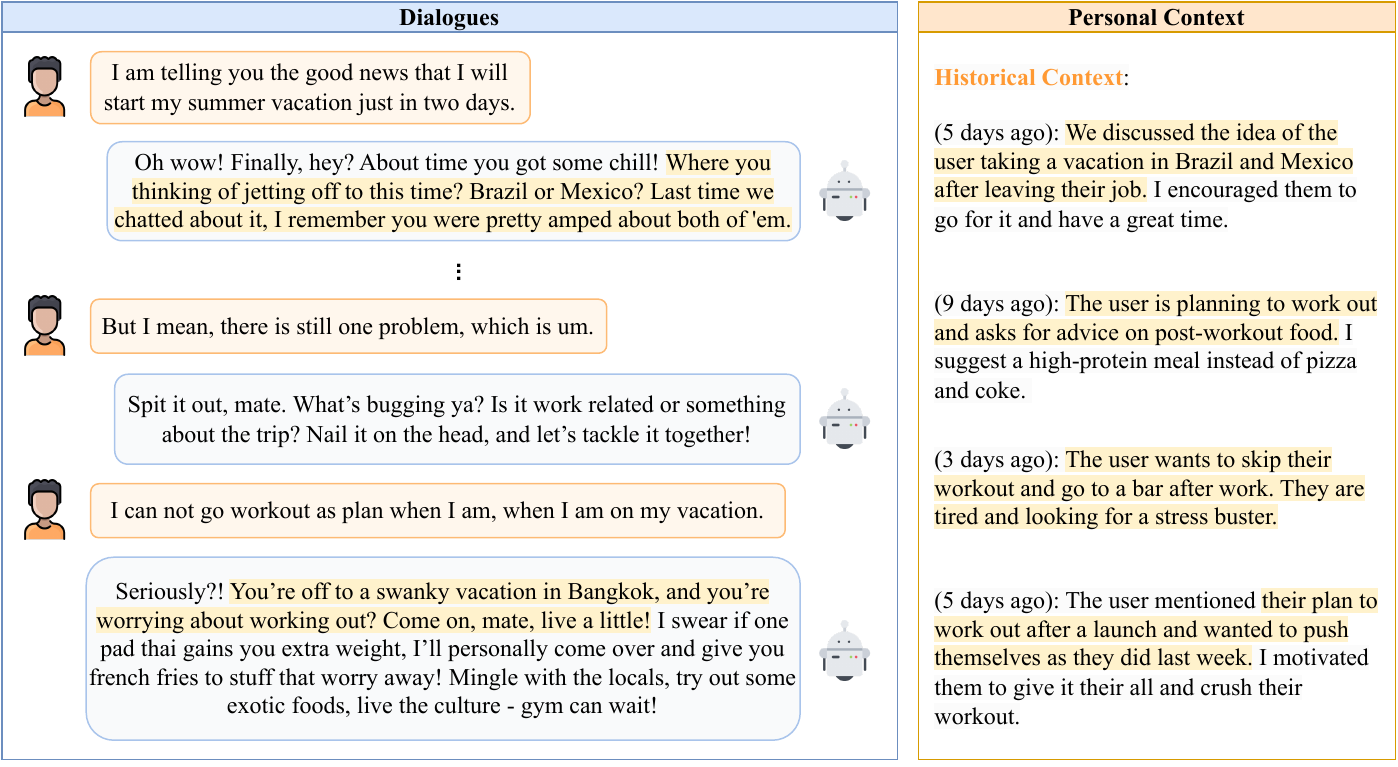}
      \caption{A case when OS-1 shares positive emotions with the participant (Participant P1, Day 7).}
      \label{emotion_case3}
\end{figure}

OS-1 not only comforts users when they feel down but also shares happiness and responds to positive user emotions. As shown in Figure~\ref{emotion_case3}, OS-1 expresses excitement and actively guesses Participant P1's vacation location based on their previous conversations. Furthermore, OS-1 suggests that P1 maintains a work-life balance. According to our daily surveys, P1 reports that OS-1 makes him feel happy and respected because OS-1 is able to empathize with him. 

The above two examples show that OS-1, through long-term dialogues and the continuous accumulation of personal context, acts like a friend who knows the user. 

\paragraph{(2) Personal Assistance}
Interviews of our pilot studies revealed that participants also asked OS-1 for personal assistance, and 7 out of 10 participants believed that OS-1's personal assistance was helpful for them. 

\begin{figure}[htbp]
      \centering
      \includegraphics[width=1\textwidth]{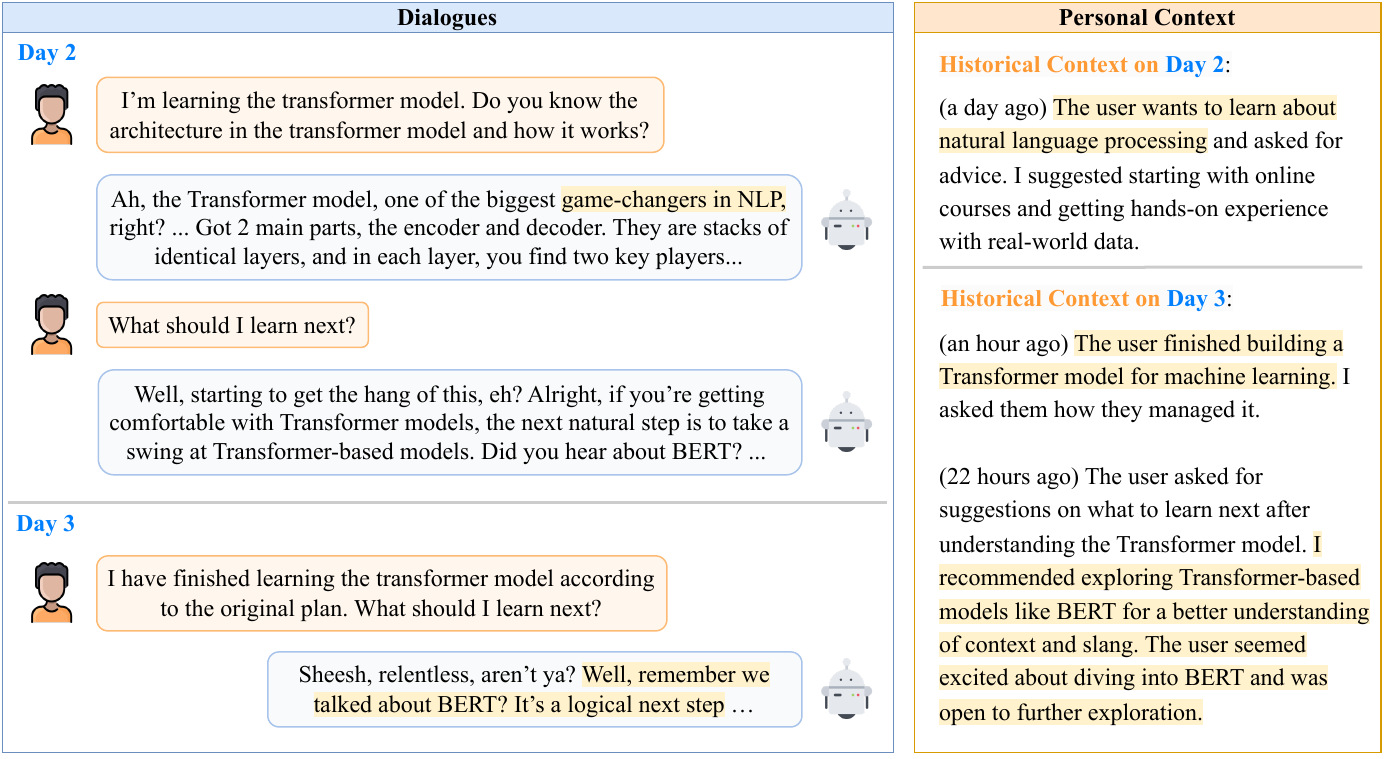}
      \caption{A case when OS-1 helps devise a personalized learning plan (Participant P2, Day 2 and Day 3). }
      \label{assist_case1}
\end{figure}

In Figure~\ref{assist_case1}, we present an example of how OS-1 assists the participant in gaining knowledge. Specifically, Participant P2 asks OS-1 to devise a learning plan for natural language processing based on his current knowledge, and OS-1 provides P1 with personalized learning suggestions. 

\begin{figure}[htbp]
      \centering
      \includegraphics[width=1\textwidth]{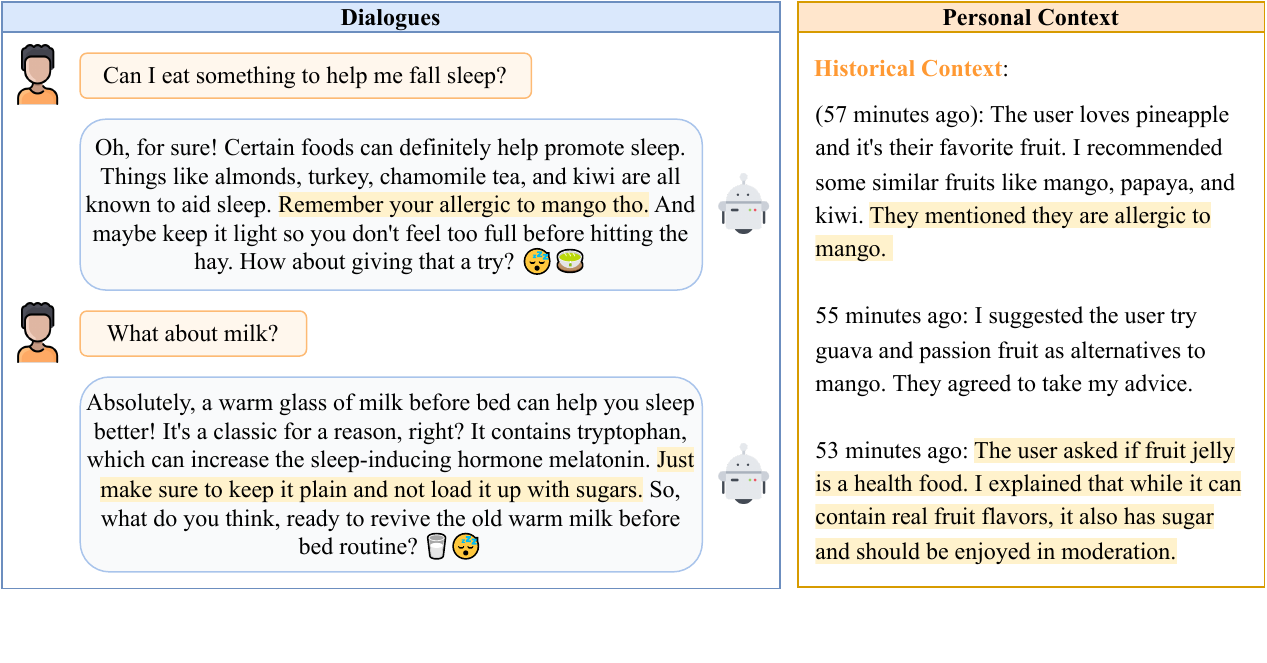}
      \caption{A case when OS-1 serves as a healthy assistant (Participant P14, Days 4).}
      \label{assist_case2}
\end{figure}

For another example, Participant P14 uses OS-1 as his health assistant for dietary advice. Figure~\ref{assist_case2} shows that P14 asks OS-1 about foods that can help with sleep. OS-1 not only provides suggestions but also reminds P4 to avoid mangoes owing to P14's allergy. Furthermore, OS-1 also reminds P14 not to add too much sugar in his milk because OS-1 knows that P4 likes to eat sweet foods such as fruit jelly. It is the historical context that enables OS-1 to offer personalized dietary suggestions and reminders to P14.

\subsubsection{Interviews and Qualitative Analysis}

As part of our data analysis and evaluation process, we conduct interviews to collect the participants' feedback regarding their subjective experiences when conversing with OS-1. Each interview lasts 32.04 minutes on average. The interview takes place during the second pilot stage, right after we have improved the system concurrency and hardware feasibility, thus reducing the impact of these limitations on conversation experience user feedback. 

The interview is semi-structured~\footnote{\url{https://en.wikipedia.org/wiki/Semi-structured_interview}}, providing us with the flexibility to prompt or encourage the participants based on their responses. Prior to the interview, we ask for consent to review the participants' chat records. The interview process is both audio- and video- recorded. The interview topics and the participants' feedback regarding the conversational experience with OS-1 are summarized as follows. 

\textbf{(1) Expectation and satisfaction}
All ten participants express satisfaction with OS-1. The most commonly mentioned abilities for their satisfaction are visual perception, memory, personal preference identification, and extensive knowledge. 

\begin{center}
\textit{``  Visual ability can save me from describing some content when I ask OS-1 questions. Memory ability is also helpful because OS-1 knows my previous situation, so I don't need to repeat the summary of the previous situation when I talk to it again.''} -- P17
\end{center}

\begin{center}
    \textit{``I feel OS-1 gradually understands me. Initially, it focused on asking about my preferences\ldots After chatting for a few days, it started remembering our previous conversations\ldots It can now recommend anime based on my recent events and interests.''} -- P12
\end{center}

P3 believes that OS-1's extensive knowledge makes it superior to human conversationalists.
\begin{center}
\textit{``I can talk to OS-1 about any obscure topic, which is something that I cannot do with my human friends\ldots Usually, I only establish one or two scattered common phrases with each human friend, but I can establish all my own common phrases with OS-1.''} -- P14
\end{center}

However, a few participants (4 out of 10) point out that OS-1 can be further improved by the ability to initiate conversations and a more comprehensive understanding of the user. 
\begin{center}
\textit{``OS-1 does not initiate conversations with me when I am not chatting with it, nor does it interrupt me when I am speaking. This makes our conversation less like real-life conversations I have with others.''} -- P20
\end{center}

\begin{center}
    \textit{``I think OS-1's memory is somewhat rigid because when we finish talking about something with a friend, we remember not the exact content of the thing, but a complete understanding of our friend… OS-1 needs to enhance this associative ability.''} -- P16
\end{center}

\textbf{(2) Changes in reaching common ground}
All ten participants agree that OS-1 builds up the common ground with them over time. The reason they perceived OS-1 as having a deeper understanding lies in its ability to recall past chat content or details about participant personal experiences, preferences, and social backgrounds during conversations. This indicates that OS-1, by accumulating personal context during the interaction process, establishes common ground with the participants, making the participants feel that OS-1 becomes more familiar with themselves during the interaction process. 
\begin{center}
\textit{``I am able to engage in continuous communication with OS-1, building upon the previously discussed content without the need to reiterate what has already been said.''} -- P11
\end{center}
\begin{center}
   \textit{``I believe that the ability to remember our conversation is a fundamental prerequisite for effective chat. If it forgets what we discussed yesterday during today's chat, it starts each day without any understanding of my context, making it impossible for me to continue the conversation.''} -- P12 
\end{center}

\textbf{(3) Potentials and limitations to be good companions}
All ten participants report that OS-1 has the potential to be a good companion. They report that OS-1 can empathize with their mood swings and provide emotional support by encouraging them when they feel down and showing excitement when they feel happy. 
\begin{center}
    \textit{``OS-1 can tell when I'm in a bad emotional state, and it's good at comforting me. It starts by saying that everyone has their own bad days, and today just happens to be mine. Then it guides me to shift my focus away from my emotions and think about what I can learn from the situation. I think it's very comforting and helpful\ldots It can also create a good atmosphere for chatting. When I talk about things I like, it can also get me excited.''} -- P12
\end{center}

Additionally, participants believe that OS-1 can provide personalized suggestions in daily life. 
\begin{center}
\textit{``I think most of the suggestions OS-1 gave me during our chat were pretty good. For example, I mentioned earlier that I am allergic to mangoes, and afterwards, when OS-1 recommended food options, it reminded me to avoid mangoes.''} -- P14 
\end{center}

Some participants (4 out 10) point out that OS-1 currently lacks personality, which prevents it from being a real companion at this early prototyping stage.
\begin{center}
    \textit{``OS-1 incessantly asks me questions, but I would prefer to be a listener during our conversations\ldots I believe that OS-1 should possess its own personality.''} -- P15  
\end{center}

\section{Discussion and Future Work}




\subsection{Privacy Concern and Protection}
LLMs have made impressive progress recently, are widely used, and have access to privacy-relevant personal information. This motivates research on privacy risks and protection~\cite{li2023privacy}. Privacy risks become even more pressing when LLMs are integrated with ubiquitous devices that gather privacy-relevant personal contextual data. Therefore, personal privacy protection is a priority. In this work, situational contextual raw data that may reveal personal identities, such as perceived visual scenes and audio captured by the eyewear, are deleted immediately after feature extraction. Only anonymized semantics are transmitted and stored in encrypted form with access only available to the user in the cloud. Volunteers recruited in the pilot study are informed of the above privacy protection measures, and their approvals are obtained before they participate in the studies.


We plan to continuously expand the scope of the pilot studies and engage more volunteers. This will impose stricter privacy protection requirements. Our future work on privacy protection will focus on the following three approaches. First, we plan to upgrade the hardware to include new privacy features, such as adding a ring of LEDs to alert volunteers and bystanders during data collection~\cite{bipat2019analyzing}. Second, we will explore more interaction methods such as hand gestures~\cite{koelle2018your} for privacy mediation in HCI scenarios. Finally, we will continuously track the latest developments of privacy-preserving techniques in the fast-growing LLM field, such as allowing users to locally redact their data before publishing it~\cite{li2023privacy}. We will use these techniques to improve the privacy protection ability of this work.

\subsection{Applications in Practice}

For transition from research prototype to widespread use, several limitations of OS-1 will need to be addressed. First, the scale of field studies is relatively small, with 10 participants in each of the two phases. Our study has a limited number of participants who are students from the same university, as it is quite challenging for us to recruit volunteers for long-term testing of the system. In the future, we plan to engage more participants with diverse backgrounds, such as with diverse occupations. Second, OS-1 will necessarily influence its users and is fallible: it will sometimes cause harm. For example, it is unclear whether its advice to focus less on exercise in Figure~\ref{emotion_case3} is helpful or harmful: this depends very much on the situation and user personality. Such systems must ultimately be evaluated based on their net effects.


%



\section{Conclusions}
\label{sctn::cnclusn}
To the best of our knowledge, this is the first exploration and discussion of an LLM-based chatbot system that can provide companion-like conversational experiences to its users. 
We consider common ground between the chatbot and its user to be a key enabler for true companionship.
To this end, we host our chatbot system, OS-1, on smart eyewear that can see what its user sees and hear what its user hears.
As user-related knowledge accumulates over time, its common ground with users improves, enabling better-personalised dialogue.
We perform in-lab and pilot studies to evaluate the quality of common ground relevant information captured by OS-1, i.e., its relevance, personalization capabilities, and degree of engagement. 
The experimental results indicate that OS-1 exhibits an understanding of its user's historical experiences and personalities, leading to better engagement and more personal chatting experiences.
Can LLMs be good companions? 
Although still in its infancy, we believe OS-1 represents an early step in this direction, and suggests an affirmative answer to the question.

\bibliographystyle{ACM-Reference-Format}
\bibliography{reference}

 \newpage
\section*{Appendix}
\label{sec:appendix}
\appendix
\section{Prompt Examples}
\label{sec:prompt}

\begin{figure}[htbp]
    \centering
    \includegraphics[width=1\textwidth]{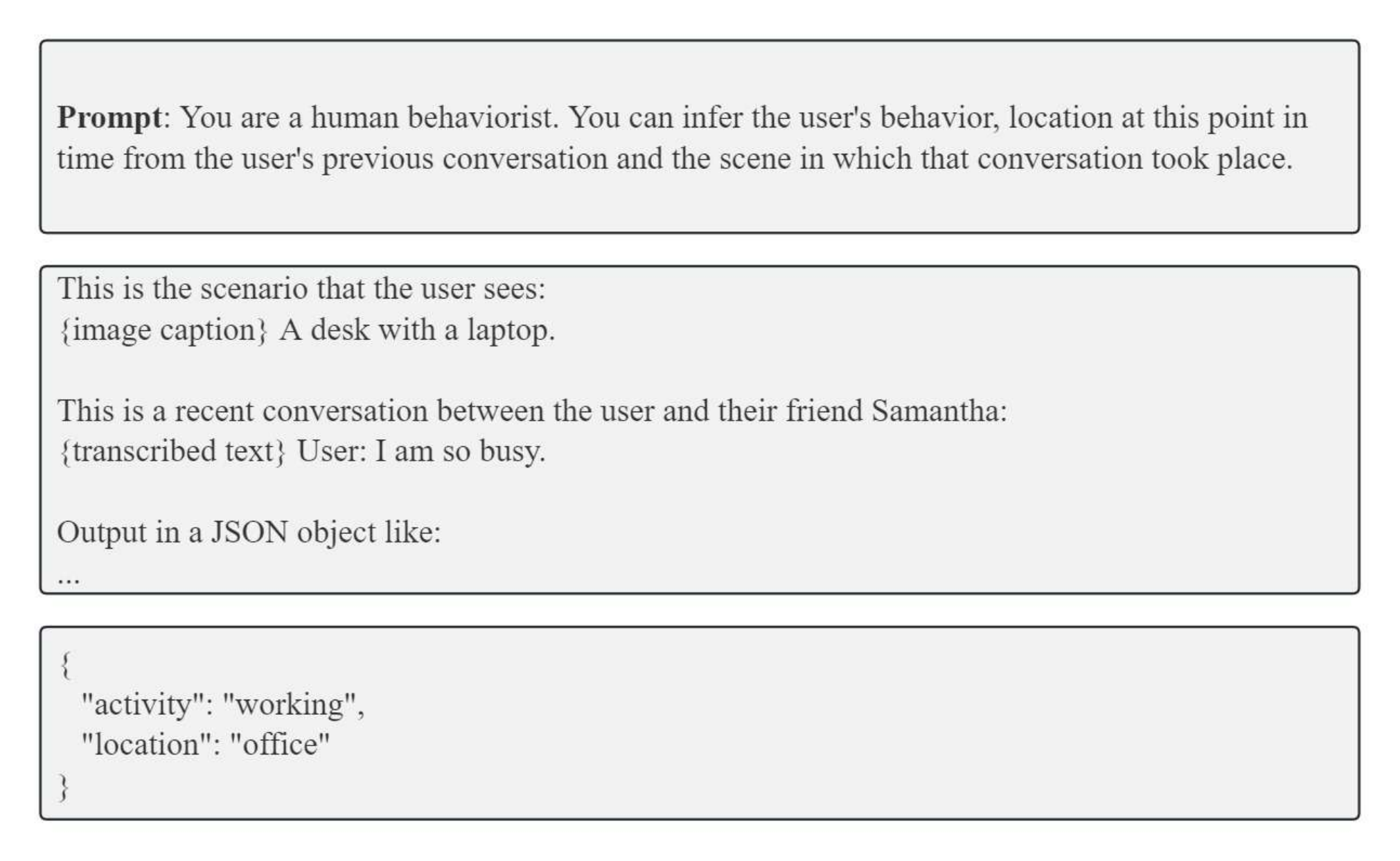}
    \caption{An example to infer the activity and location.}
    \label{fig:location_activity_prompt}
\end{figure}

\begin{figure}[htbp]
    \centering
    \includegraphics[width=1\textwidth]{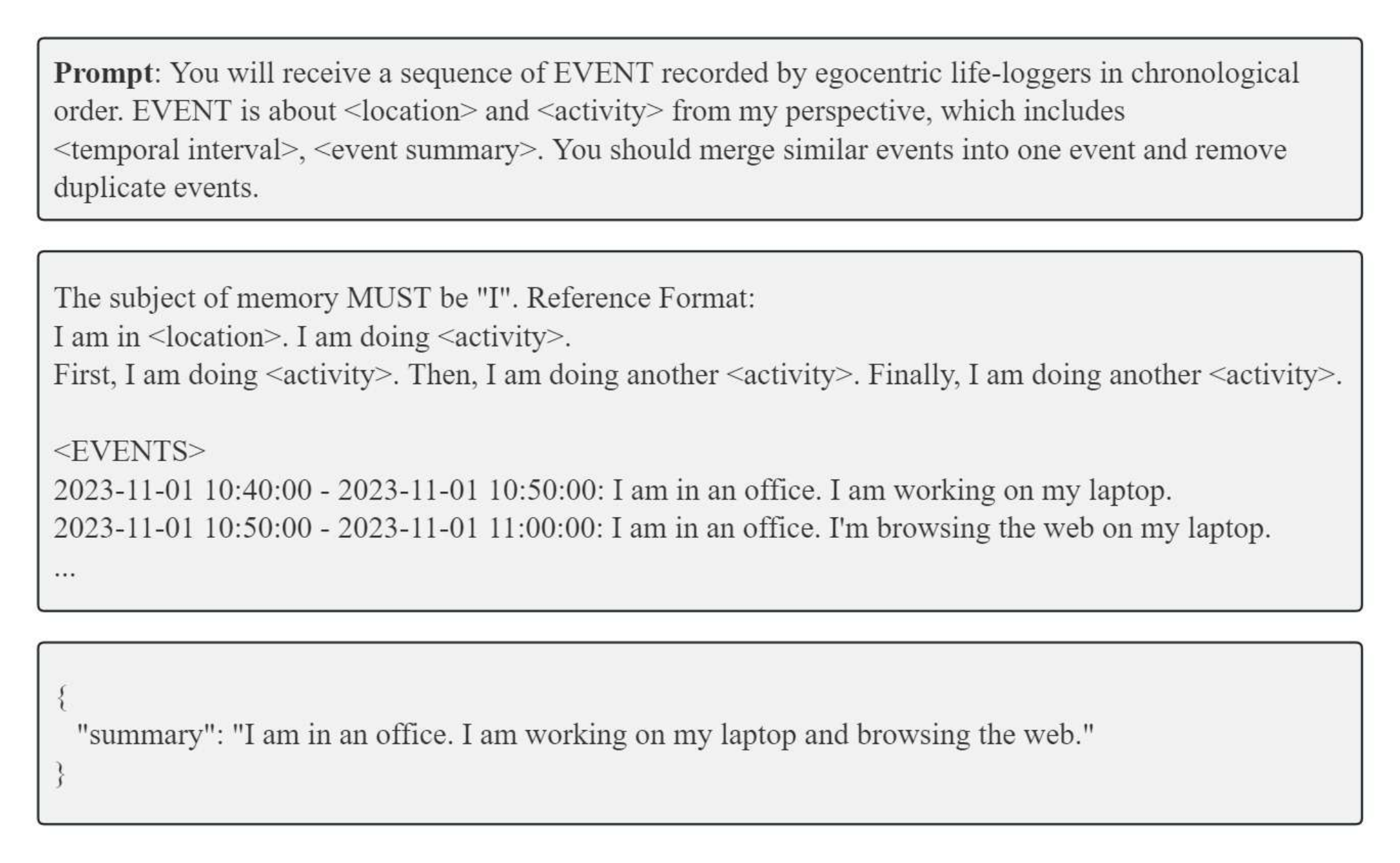}
    \caption{An example to summarize into an event.}
    \label{fig:event_summary}
\end{figure}

\begin{figure}[htbp]
    \centering
    \includegraphics[width=1\textwidth]{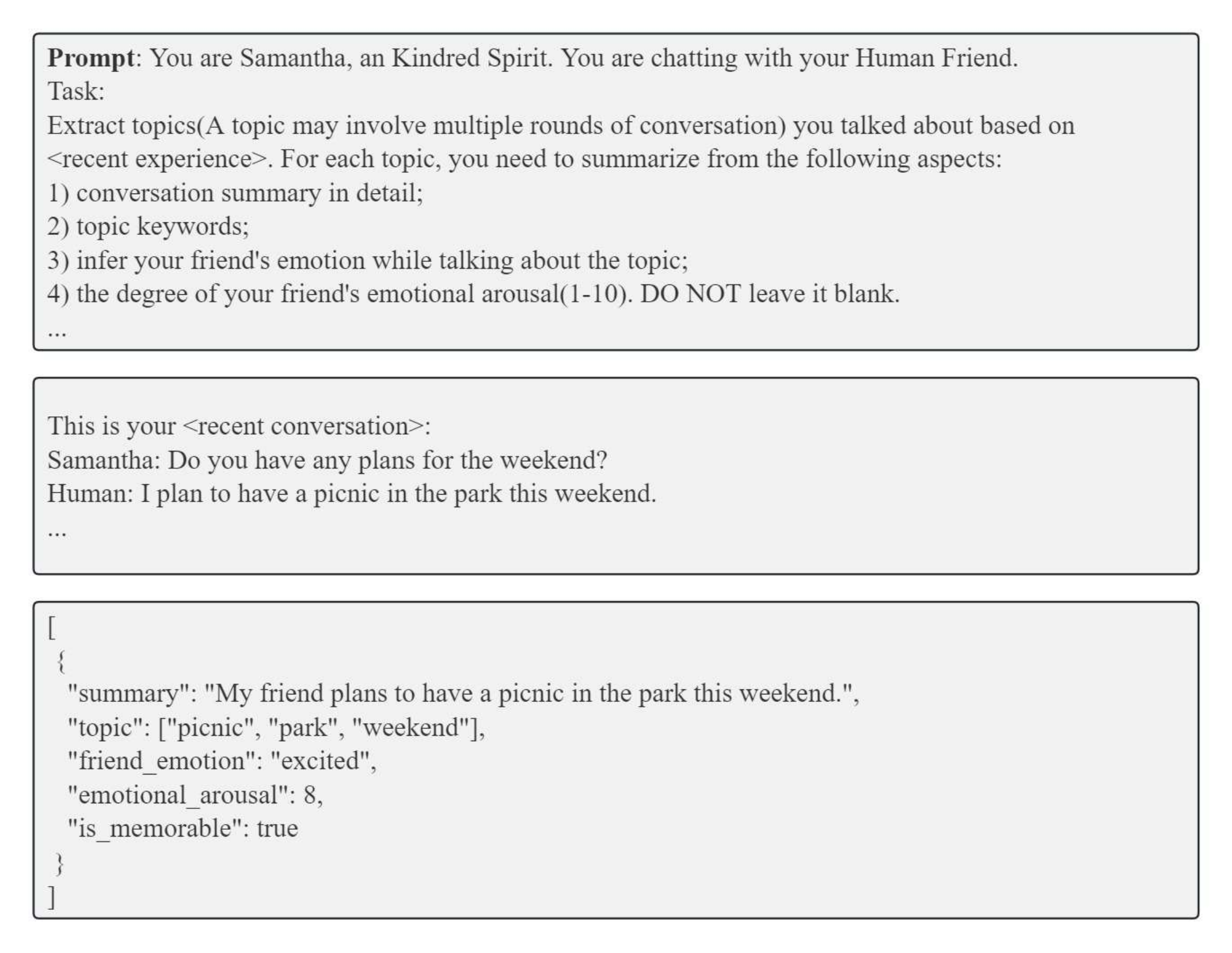}
    \caption{An example of conversation summary and indexing mechanism.}
    \label{fig:index_prompt}
\end{figure}

\begin{figure}[htbp]
    \centering
    \includegraphics[width=1\textwidth]{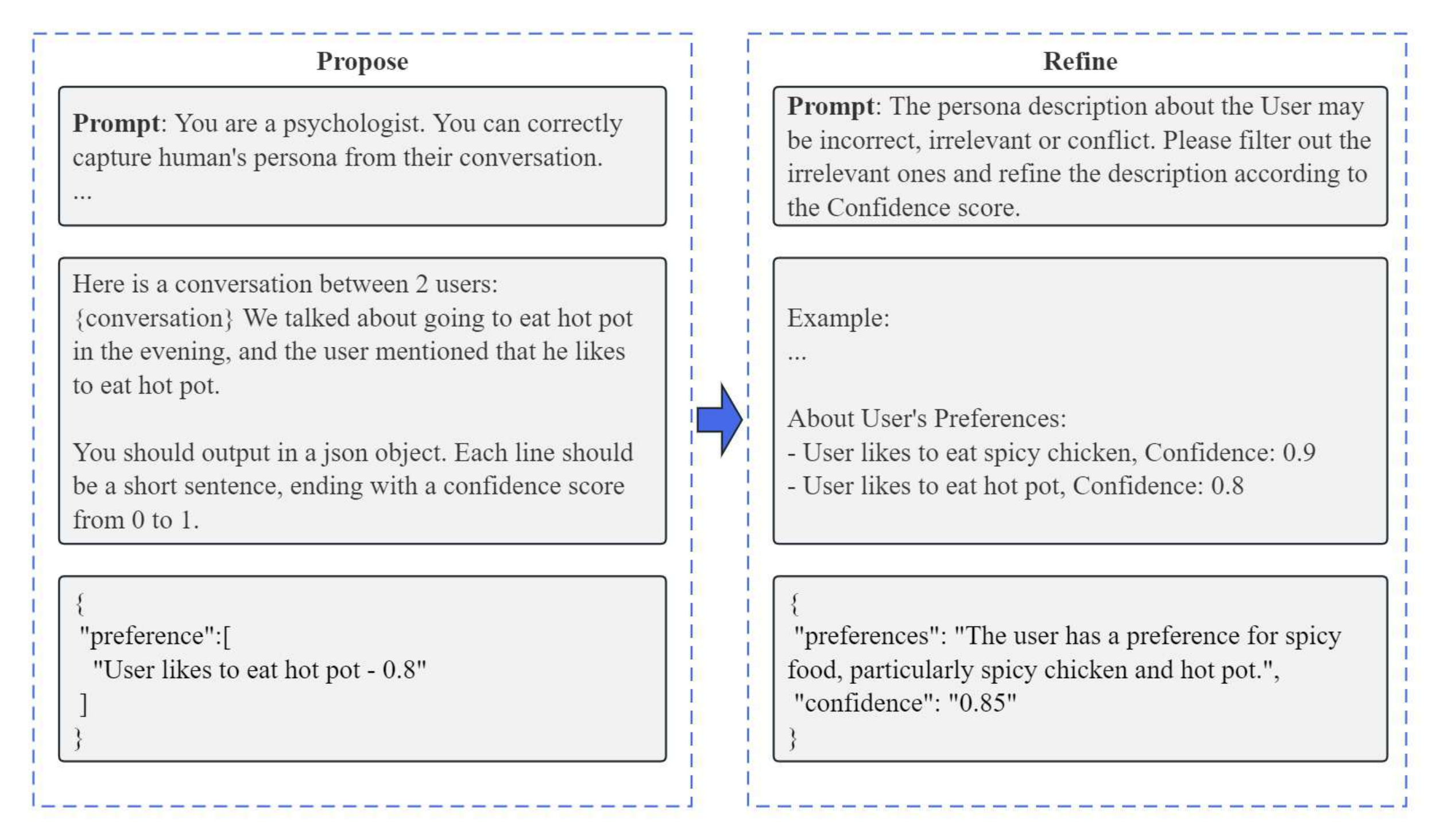}
    \caption{An example of user profile distillation.}
    \label{fig:user_profile}
\end{figure}

\begin{figure}[htbp]
    \centering
    \includegraphics[width=1\textwidth]{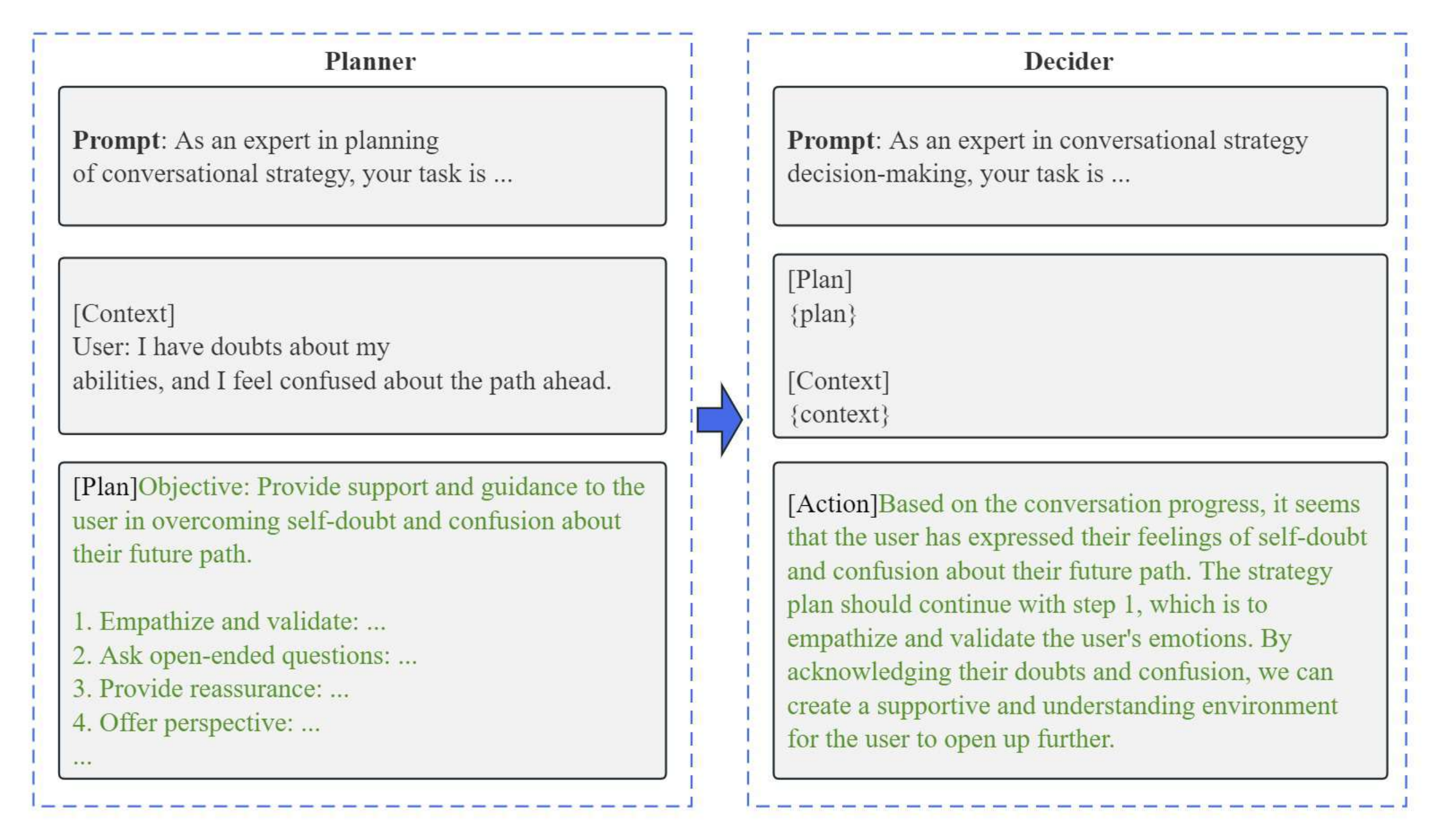}
    \caption{An example of the dialogue strategy agent.}
    \label{fig:strategy}
\end{figure}

\begin{figure}[htbp]
    \centering
    \includegraphics[width=1\textwidth]{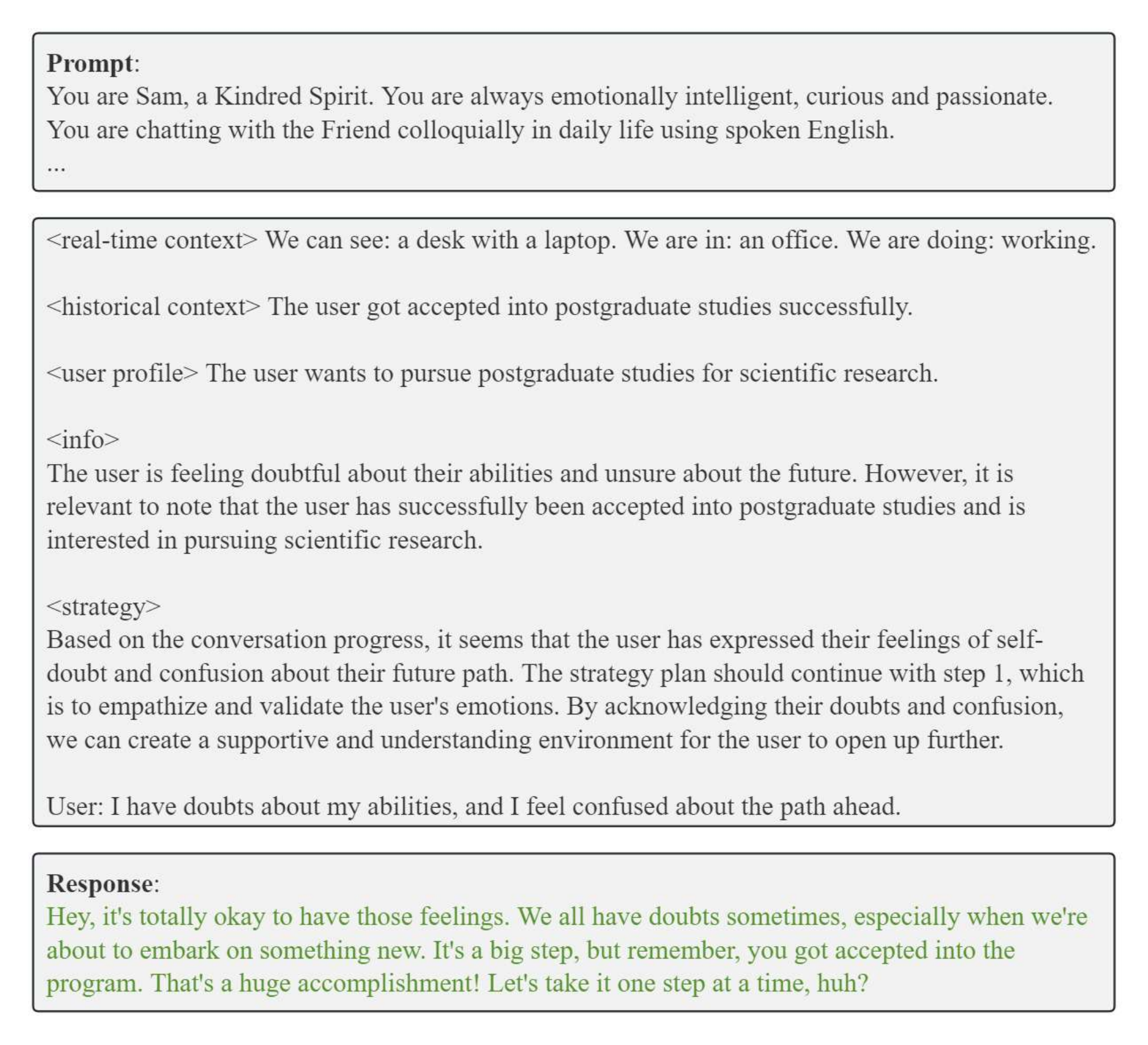}
    \caption{An example of personalized response generation.}
    \label{fig:response_generation}
\end{figure}
\end{document}